\documentclass[numberedappendix,twocolumn,twocolappendix]{openjournal}

\usepackage{latexsym}
\usepackage{graphicx}
\usepackage{amssymb}
\usepackage{longtable}
\usepackage{epsf}
\usepackage{amsmath}
\usepackage{graphicx}
\usepackage{textcomp}
\usepackage{gensymb}

\usepackage{hyperref}
\hypersetup{colorlinks=true,linkcolor=blue,citecolor=blue,filecolor=blue,urlcolor=blue}
\usepackage{cleveref}
\usepackage{bm}		

\usepackage{natbib}

\newcommand{\msun}{{\rm M}_{\odot}}

\newcommand{\beq}{\begin{eqnarray}}
\newcommand{\eeq}{\end{eqnarray}}
\newcommand{\ben}{\begin{itemize}}
\newcommand{\een}{\end{itemize}}

\newcommand{\mpeak}{M_{\rm peak}}
\newcommand{\mhalo}{M_{\rm halo}}
\newcommand{\dmhdtt}{{\rm d}M_{\rm peak}/{\rm d}t}
\newcommand{\dmhdt}{{\rm d}M_{\rm peak}/{\rm d}t}
\newcommand{\mzero}{M_{0}}

\newcommand{\tauc}{\tau_{\rm c}}
\newcommand{\aearly}{\alpha_{\rm early}}
\newcommand{\alate}{\alpha_{\rm late}}

\newcommand{\lge}{\beta_{\rm e}}
\newcommand{\lgl}{\beta_{\ell}}
\newcommand{\lgtc}{x_{0}}

\newcommand{\flate}{F_{\rm late}}
\newcommand{\fbarlate}{\bar{F}_{\rm late}}
\newcommand{\mui}{\mu_{\rm i}}
\newcommand{\mubari}{\bar{\mu}_{\rm i}}

\newcommand{\sigi}{\Sigma_{\rm i}}

\newcommand{\sigbari}{\bar{\Sigma}_{\rm i}}

\newcommand{\tform}{t_{\rm form}}
\newcommand{\tperc}[1]{t_{#1\%}}

\usepackage{xspace}
\newcommand{\tng}{IllustrisTNG\xspace}
\newcommand{\dmah}{Diffmah\xspace}
\newcommand{\dmahpop}{DiffmahPop\xspace}

\newcommand{\mahpdf}{P(\mpeak(t)\vert\mzero)}
\newcommand{\dmhdtpdf}{P(\dmhdtt\vert\mzero)}
\newcommand{\parampdf}{P(\aearly, \alate, \tauc\vert\mzero)}

\begin{document}

\title{A Differentiable Model of the Assembly of Individual and Populations of Dark Matter Halos}
\shorttitle{A Differentiable Model of Halo Assembly}

\author{Andrew P. Hearin$^{1,\star}$}
\author{Jon\'{a}s Chaves-Montero$^{1,2}$}
\author{Matthew R. Becker$^1$}
\author{Alex Alarcon$^1$}

\affiliation{$^1$HEP Division, Argonne National Laboratory, 9700 South Cass Avenue, Lemont, IL 60439, USA}
\affiliation{$^2$Donostia International Physics Centre, Paseo Manuel de Lardizabal 4, 20018 Donostia-San Sebastian, Spain}
\thanks{$^{\star}$E-mail:ahearin@anl.gov}

\shortauthors{Hearin et al.}

\begin{abstract}
We present a new empirical model for the mass assembly of dark matter halos. We approximate the growth of individual halos as a simple power-law function of time, where the power-law index smoothly decreases as the halo transitions from the fast-accretion regime at early times, to the slow-accretion regime at late times. Using large samples of halo merger trees taken from high-resolution cosmological simulations, we demonstrate that our 3-parameter model, \dmah, can approximate halo growth with a typical  accuracy of 0.1 dex for $t\gtrsim1\,{\rm Gyr}$ for all halos of present-day mass $M_{\rm halo}\gtrsim10^{11}\msun,$ including subhalos and host halos in gravity-only simulations, as well as in the \tng hydrodynamical simulation. We additionally present a new model for the assembly of halo populations, \dmahpop, which not only reproduces average mass growth across time, but also faithfully captures the {\em diversity} with which halos assemble their mass. Our python implementation is based on the autodiff library JAX, and so our model self-consistently captures the mean and variance of halo mass accretion rate across cosmic time. We show that the connection between halo assembly and the large-scale density field, known as {\em halo assembly bias}, is accurately captured by \dmah, and that residual errors in our approximations to halo assembly history exhibit a negligible residual correlation with the density field. Our publicly available source code can be used to generate Monte Carlo realizations of cosmologically representative halo histories; our differentiable implementation facilitates the incorporation of our model into existing analytical halo model frameworks.
\end{abstract}

\maketitle

\vspace{1cm}

\twocolumngrid

\section{Introduction}
\label{sec:intro}

In the standard cosmological model, the matter content of the Universe is dominated by Cold Dark Matter (CDM), and gravitationally self-bound objects referred to as dark matter halos are the fundamental building blocks of structure formation \citep[see][for a comprehensive review]{mo_vdb_white_2010}. Dark matter halos are the natural sites of galaxy formation \citep{white_rees_1978,blumenthal_etal_1984}, and so a detailed understanding of the buildup and evolution of halos is a key ingredient of any theory of structure growth.

The basic physical picture of halo evolution is now relatively well understood. At early times, halo assembly is characterized by a ``fast-accretion phase'' in which the growth of halo mass is rapid and major mergers are common; mass accretion rates diminish considerably at later times in a ``slow-accretion phase'', during which time major mergers are comparatively rare \citep{bullock_etal01,wechsler_etal2002, tasitsiomi_kravtsov_2004}. The mass density profile of dark matter halos is well-described by a double power law known as the NFW profile \citep{nfw_1997}, an approximation that remains reasonably accurate across most of cosmic time \citep[although see][for shortcomings of this approximation]{ludlow_etal13}. Although the concentration parameter that defines the NFW profile exhibits a well-known dependence upon total mass \citep[e.g.,][]{diemer_kravtsov_2015,child_habib_etal_2018}, all halos in the fast-accretion phase tend to have similar values of concentration $c\approx3-4$ \citep{zhao_mo_etal03}, and as halos transition to the slow-accretion phase, their concentrations steadily increase as they predominantly pile up mass onto their outskirts \citep[see][for a detailed physical picture of the relationship between halo assembly and NFW concentration]{wang_zentner_etal_2020}.

Many additional characteristics of the internal structure of dark matter halos are closely related to their assembly histories, including substructure abundance \citep{gao_etal04,wechsler_etal2006,giocoli_tormen_sheth_vdb2010}, the ``splashback'' feature in the outer profile \citep{diemer_kravtsov_2014}, and ellipsoidal shape \citep{chen_mo_2020,lau_hearin_2021}. The growth history of a dark matter halo is also tightly connected to the larger-scale environment in which it evolves, a phenomenon known as halo assembly bias \citep[e.g.,][]{gao_springel_white_2005,mao_zentner_wechsler_2018,mansfield_kravtsov_2020}.

The mass assembly history of dark matter halos is widely used in theoretical models of galaxy formation as a fundamental quantity regulating the star formation rate of the halo's resident galaxy, including early formulations of traditional semi-analytic modeling \citep[][SAMs, hereafter]{kauffmann_white_Guiderdoni_1993,baugh_cole_frenk_1996,avila_reese_1998}, practically all contemporary SAMs \citep[e.g.,][]{bower_durham_sam_2006,somerville_etal_2008,henriques_etal_2015,lagos_shark_2018}, simple empirical models \citep{watson_hearin_age_matching3,behroozi_silk_2015}, and recent empirical approaches \citep{becker_smad_2015,moster_emerge1, behroozi_etal19}. Strong motivation for treating halo assembly as the backbone of galaxy formation models also comes directly from hydrodynamical simulations, which show that the accretion rate of {\em baryonic} mass closely tracks that of dark matter \citep{wetzel_nagai_2015}.

The assembly of a halo is fully characterized by its {\em merger tree,} which describes the mass and accretion time of the ``progenitors'' that merged into the main halo over time, as well as the progenitors of those progenitors, etc. The most common theoretical technique for modeling halo merger trees is the (extended) Press-Schechter formalism \citep[][EPS, hereafter]{press_schechter_1974,bond_etal91, bower_1991, lacey_cole_93}. The EPS framework yields considerable insight into many aspects of the structure and buildup of dark matter halos \citep[e.g.,][]{somerville_kolatt_1999,nusser_sheth1999,cole_frenk_parkinson_2008}, and has been shown to yield reasonably accurate predictions for the average growth of halos over time \citep{van_den_bosch2002}, the mass function of the progenitor halos \citep{parkinson_cole_helly_2008}, and the statistical distribution of halo formation times \citep{li_etal07,giocoli_etal12}. Merger trees predicted by EPS furthermore enable suitably implemented SAMs to generate predictions for galaxy evolution without the need for high-resolution N-body simulations \citep[e.g.,][]{somerville_etal_2008,benson_galacticus_2012}. We refer the reader to \citet{zentner_eps_2007} for a pedagogical overview of EPS, and to \citet{,jiang_vdb_2014b} for a comparison of modern implementations of EPS predictions of merger trees.

Empirically formulated fitting functions calibrated against cosmological simulations offer a simpler route to describing the evolutionary history of halo mass. The one-parameter exponential model developed in \citet{wechsler_etal2002} is one of the first of such models, and a number of improvements have followed this early empirical approach that capture richer phenomena with greater accuracy. For example, the fitting formula in \citet{tasitsiomi_kravtsov_2004} is a two-parameter generalization of the exponential growth model with improved performance for a greater diversity of cluster-mass halos. In \citet{fakhouri_ma_boylan_kolchin_2010}, the authors present a simple and accurate fitting formula for $\langle dM_{\rm halo}(t)/dt\vert\mzero\rangle,$ the average mass accretion rate of halos as a function of cosmic time and present-day mass $\mzero,$ and they furthermore find that they can use their model to accurately predict $\langle M_{\rm halo}(t)\vert\mzero\rangle$ by numerically integrating their parameterized mass accretion rate histories. In \citet{mcbride_fakhouri_ma_2009}, the authors introduce a different two-parameter fitting function for individual halo mass growth, and also include additional modeling ingredients for the dependence of how the fitting parameters depend on present-day halo mass.

The program to develop empirical approximations to dark matter halo assembly is faced with a number of challenges. Each of the efforts summarized above were conducted at fixed cosmology, and also for gravity-only N-body simulations, and the robustness of these findings to effects of cosmology and hydrodynamics remains relatively coarsely characterized \citep[although see][discussed further in \S\ref{sec:discussion} below]{vdb_etal14}. Efforts to improve flexibility beyond the one-parameter exponential model also face a significant difficulty of ensuring physically consistent behavior across the full range of mass and redshift; for example, as pointed out in Appendix H of \citet{behroozi_etal13}, predictions of the model presented in \citet{mcbride_fakhouri_ma_2009} exhibit unphysical behavior at high redshift due to the crossing of average assembly histories of halos of different present-day mass. Of special concern to the present paper, empirical fitting functions historically only reproduce {\em average} halo growth across time, but do not enjoy capability to predict the {\em diversity} of halo assembly, a shortcoming made especially glaring through comparison to EPS-based techniques.

In this paper, we seek to improve upon this situation by developing an empirical approach to modeling the growth of both individual halos, as well as the assembly of cosmologically representative populations. In \S\ref{sec:individual_halos}, we will introduce a new parameterization for individual halo growth, \dmah, validating that our formulation is sufficiently flexible to capture the assembly of halos in both the gravity-only and full-physics hydro simulations described in \S\ref{sec:sims}. In \S\ref{sec:halo_populations}, we outline \dmahpop, our statistical model for the assembly of halo populations, relegating exposition on mathematical and computational details to the appendices. We discuss our results and potential applications in \S\ref{sec:discussion}, and conclude in \S\ref{sec:summary} with a brief summary of our primary results.



\section{Simulations and Merger Trees}
\label{sec:sims}

We have built our model for the mass assembly history (MAH hereafter) of dark matter halos to approximate results taken from the merger trees of halos identified in both gravity-only and hydrodynamical simulations. For our gravity-only simulations, we use a combination of the Bolshoi Planck simulation \citep[BPL,][]{klypin_etal11} and MultiDark Planck 2 \citep[MDPL2,][]{klypin_etal16}. The BPL simulation was carried out using the ART code \citep{kravtsov_etal97_art} by evolving $2048^3$ dark-matter particles of mass $m_{\rm p}=1.55\times10^{8}\msun$ on a simulation box of $250\,{\rm Mpc}$ on a side; the MDPL2 simulation was carried out using the L-GADGET-2 code \citep{springel_2005_gadget2} with $3840^3$ dark-matter particles of mass $m_{\rm p}=1.51\times10^{9}M_{\odot}$ on a simulation box of $1000\,{\rm Mpc}$ on a side; both simulations were run under cosmological parameters closely matching \citet{planck14b}. For both BPL and MDPL2, we used publicly available\footnote{\url{https://www.peterbehroozi.com/data.html}} merger trees that were identified with Rockstar and ConsistentTrees \citep{behroozi_etal13_rockstar, behroozi_etal13_consistent_trees, rodriguez_puebla_etal16}.

We additionally studied the assembly of halos in the \tng hydrodynamical simulations. \tng is a suite of hydro simulations that models the joint evolution of dark matter, gas, stars, and supermassive black holes, and includes treatment of radiative gas cooling, star formation, galactic winds, and AGN feedback \citep{Weinberger2017, Pillepich2018a}. We use publicly available data from the largest hydrodynamical simulation of the suite, TNG300-1 \citep{Nelson2018a}. The TNG300-1 simulation was carried out using the moving-mesh code {\sc Arepo} \citep{Springel2010} $2500^3$ with gas tracers together with the same number of dark matter particles in a simulation box of $205$ Mpc on a side under a cosmology very similar to \citet{planck14b}. For TNG300-1 the corresponding mass resolution is $4.1\times10^7\msun$ and $7.7\times10^6\msun$ for dark matter and gas, respectively. Halos and subhalos in \tng were identified with the {\tt SUBFIND} algorithm \citep{springel_etal2001_subfind}, and the merger trees we use were constructed with {\tt SUBLINK} \citep{rodriguez_gomez_etal15_sublink}.

For some of the results in the paper, we calculate cross-correlation functions between simulated halos and the density field, using a random downsampling of particles from the appropriate snapshot. For BPL we use a random downsampling of dark matter particles provided by {\tt halotools} \citep{hearin_etal17_halotools}.
For \tng, we use a random downsampling of TNG300-1 particles at $z=0$ kindly provided by Benedikt Diemer.

All results in the main body of the paper pertain to the assembly histories of present-day host halos (i.e., {\tt upid=-1} for Rockstar, and the ``main halo'' for {\tt SUBFIND}). Comparable results for the case of subhalos can be found in Appendix~\ref{appendix:subs}. Throughout the paper, including the present section, values of mass and distance are quoted assuming $h=1.$ For example, when writing $\mpeak=10^{12}\msun,$ we suppress the $\msun/h$ notation and write the units as $\msun.$

\begin{figure*}
\begin{centering}
\includegraphics[width=14cm]{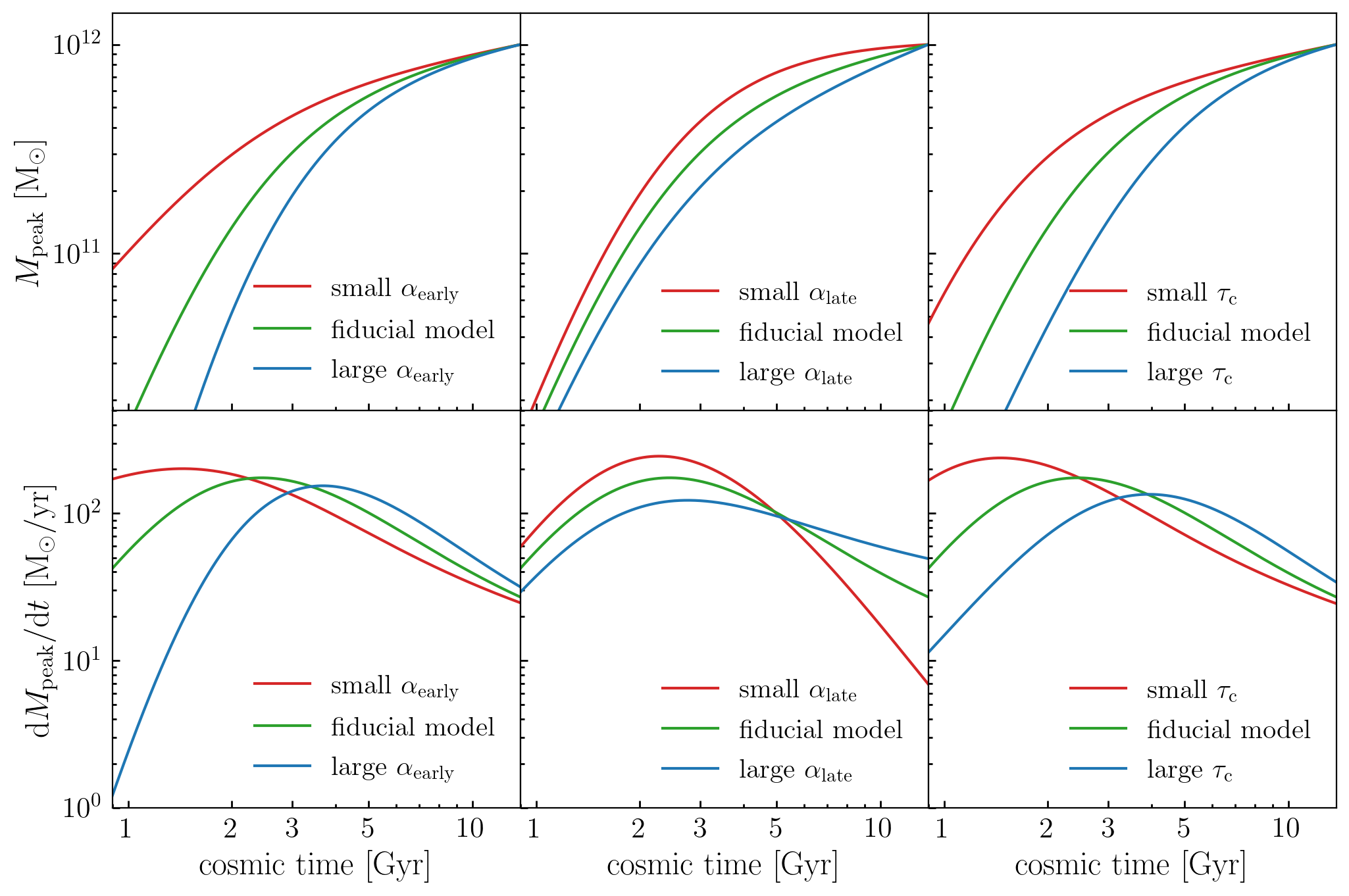}
\caption{{\bf Physical interpretation of the parameters of the \dmah model for individual halo assembly}. Each panel shows the assembly history of a dark matter halo with present-day mass $\mzero=10^{12}\msun;$ the top row of panels shows the history of cumulative peak mass, $\mpeak(t);$ the bottom row of panels shows the history of mass accretion rate, $\dmhdtt.$ Our model approximates halo assembly as a power-law function of time with rolling index, $\mpeak(t)\propto t^{\alpha(t)},$ as in Eq.~\ref{eq:individual_mah}. The left columns show the influence of $\aearly,$ the power-law index at early times; the middle columns show the influence of $\alate,$ the power-law index at late times; the right columns show $\tauc,$ which controls the transition time between the fast- and slow-accretion regimes. Smaller values of each parameter correspond to halos with earlier formation times.}
\label{fig:diffmah_parameter_variations}
\end{centering}
\end{figure*}



\section{Modeling Individual Halo Assembly with Diffmah}
\label{sec:individual_halos}

\begin{figure}
\includegraphics[width=7.5cm]{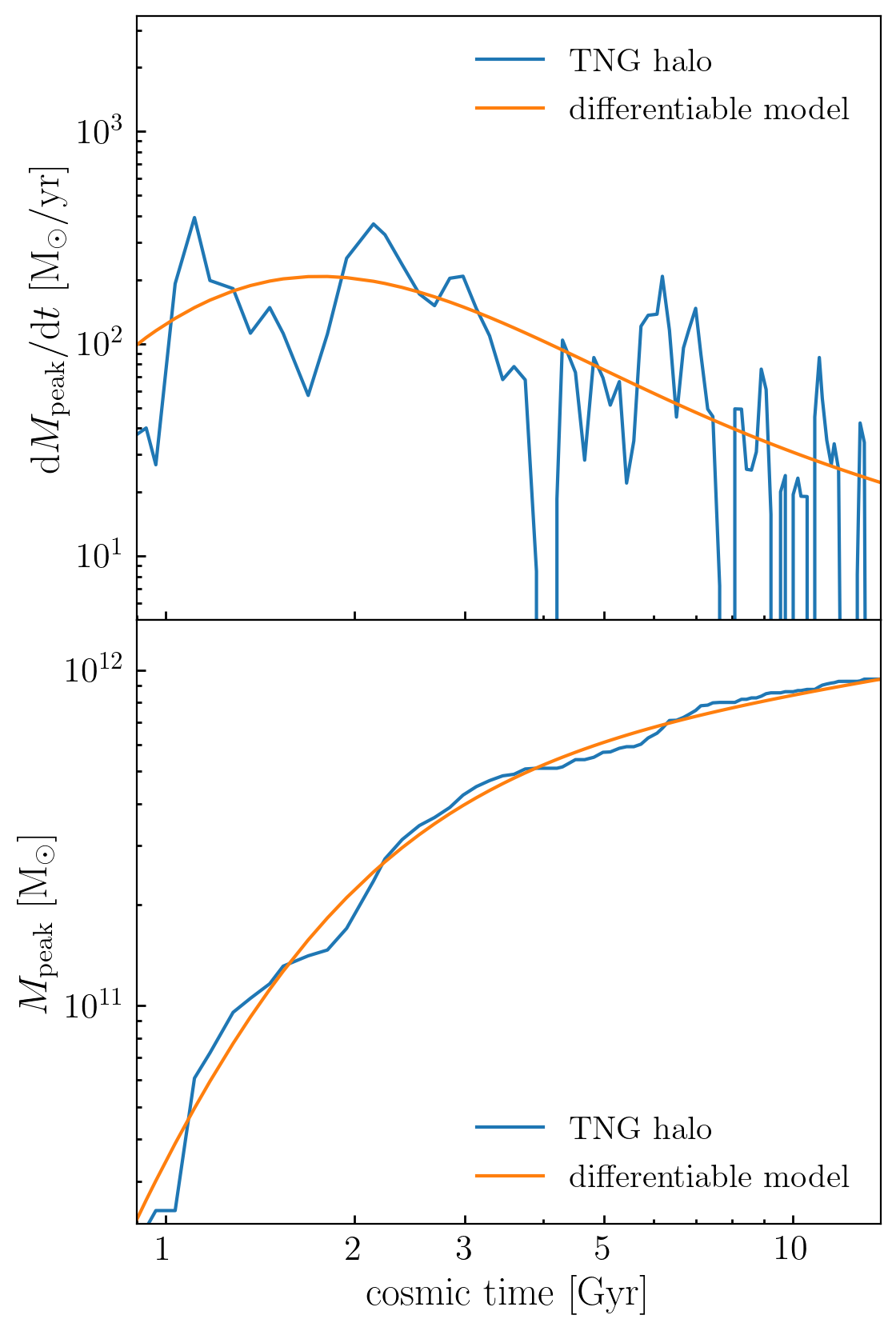}
\caption{{\bf Example fit to individual halo growth}. The blue curve in each panel shows the assembly history of the same dark matter halo taken directly from the merger trees of the \tng simulation. The top panel shows the history of the halo's mass accretion rate, and the bottom panel shows the history of its cumulative peak mass. In each panel, the orange curve shows the approximate halo history based on our differentiable model.}
\label{fig:example_mah}
\end{figure}

In this section, we present results for \dmah, our model of the assembly of individual dark matter halos. In \S\ref{subsec:individual_halo_model} we describe the functional form of the model and its differentiable implementation, and in \S\ref{subsec:individual_halo_results} we show the accuracy with which the model can approximate the formation history of simulated halos.

\subsection{\dmah Model formulation}
\label{subsec:individual_halo_model}

The \dmah model for the growth of individual halos is designed to capture the evolution of {\em cumulative peak} halo mass, $\mpeak(t),$ defined as the largest mass the main progenitor halo has ever attained up until the time $t.$ Thus as a matter of definition, $\mpeak(t)$ is a non-decreasing function for all $t,$ so that $\mpeak(t)$ will remain constant if a halo experiences a period of mass loss.

We model $\mpeak(t)$ with a power-law function of cosmic time with a rolling index,
\beq
\label{eq:individual_mah}
\mpeak(t) = \mzero(t/t_0)^{\alpha(t)},
\eeq
where $t_0$ is the present-day age of the universe, $\mzero\equiv\mpeak(t_0),$ and $\alpha(t)$ is given by a sigmoid function defined as follows:
\beq
\label{eq:sigmoid}
\alpha(t; \tau_{\rm c}, k, \aearly, \alate)\equiv \aearly + \frac{\alate-\aearly}{1 + \exp(-k(t-\tauc))}.
\eeq
The parameters $\aearly$ and $\alate$ define the asymptotic value of the power-law index at early and late times, respectively, and $\tauc$ is the transition time between the early- and late-time indices. The parameter $k$ controls the speed of the transition between the two regimes; as described in Appendix~\ref{appendix:individual_growth}, for all results in the paper we hold $k$ fixed to a constant value of $3.5.$

In Figure~\ref{fig:diffmah_parameter_variations}, we give a visual illustration of the physical interpretation of the three free parameters of our model: $\aearly,$ $\alate,$ and $\tauc.$ Each of the six panels shows the assembly history of halos with present-day mass $\mzero=10^{12}\msun,$ with variations of the MAH parameters color-coded according to the legend. For each of the three MAH parameters, $\aearly,$ $\alate,$ and $\tauc,$ smaller values of the parameter corresponds to a halo that was accreting mass more rapidly at early times and more slowly today; thus a halo with a smaller value of a MAH parameter has an earlier formation time relative to a halo with a larger value of that parameter. For curves labeled ``fiducial model'', we have $(\aearly, \alate, \tauc)=(2.5, 0.3, 1.25 {\rm [Gyr]});$ for parameters set to their smaller values, we have $(\aearly, \alate, \tauc)=(1.25, 0.05, 0.6 {\rm [Gyr]});$ for larger values, we have $(\aearly, \alate, \tauc)=(5.0, 0.6, 2.5 {\rm [Gyr]}).$ The top row of panels shows the history of cumulative mass, $\mpeak(t),$ which is parameterized directly according to Eq.~\ref{eq:individual_mah}. The bottom row of panels shows $\dmhdt,$ the history of mass accretion rate. While $\dmhdtt$ is analytically calculable by symbolically differentiating Eq.~\ref{eq:individual_mah}, we have computed each curve in the bottom panels of Figure~\ref{fig:diffmah_parameter_variations} via automatic differentiation \citep[autodiff, see][for a recent review]{baydin_etal15}, as our source code is implemented based on the JAX autodiff library \citep{jax2018github}.

\subsection{\dmah Model performance}
\label{subsec:individual_halo_results}

\begin{figure}
\includegraphics[width=7.5cm]{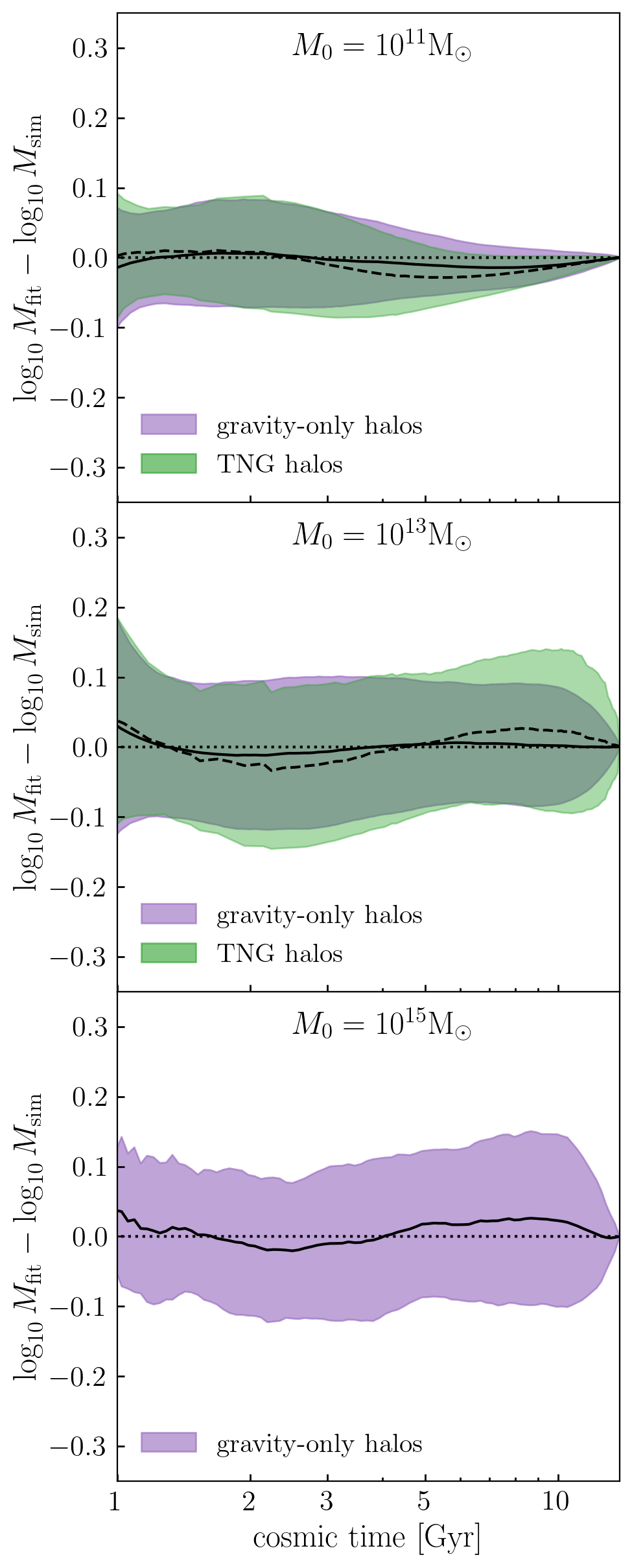}
\caption{{\bf Residuals of \dmah fits to individual halos}. Using on a large collection of fits to the assembly histories of simulated halos, each panel shows the logarithmic difference between the simulated and approximate values of cumulative peak mass, $\mpeak(t),$ as a function of time; results for samples of halos of different present-day mass are shown in different panels as indicated by the in-panel annotation. The black curve in each panel shows the average residual difference, and the shaded band shows the $1\sigma$ scatter; residuals for gravity-only simulations are shown with the solid black curve and purple shaded region; residuals for \tng are shown with the dashed black curve and green shaded region. The figure shows that the rolling power-law model defined by Eq.~\ref{eq:individual_mah} gives an unbiased approximation to halo mass assembly with a typical error of 0.1 dex for $t\gtrsim1$ Gyr. }
\label{fig:individual_residuals}
\end{figure}

In the previous section, we illustrated the basic features of the \dmah model for dark matter halo assembly. Here we show the accuracy with which simulated halo histories can be approximated by \dmah. When fitting the mass growth of an individual dark matter halo with the fitting function in Equation \ref{eq:individual_mah}, we allow the parameters $\aearly,$ $\alate,$ and $\tau_{\rm c}$ to vary freely, and use gradient descent to minimize the logarithmic difference between the simulated and predicted values of $\mpeak(t),$ with $k$ held fixed to a constant value of $3.5$ for all halos, and $\mzero$ held fixed to $\mpeak(t_0).$ We refer the reader to Appendix~\ref{appendix:individual_growth} for a detailed description of our optimization technique.

\subsubsection{Capturing individual halo growth}
\label{subsec:individual_halo_residuals}

Figure~\ref{fig:example_mah} shows a typical example of the ability of \dmah to approximate individual dark matter halo growth. Both panels show the assembly history of the same dark matter halo taken from the \tng simulation; the top panel shows the history of the halo's mass accretion rate, and the bottom panel shows the history of its cumulative peak mass. The blue curve in each panel shows the assembly history of the halo as taken directly from the simulated merger tree, and the orange curve shows the differentiable approximation. As described in Appendix~\ref{appendix:individual_growth}, our differentiable approximation comes from fitting $\mpeak(t)$ of the simulated halo, and our approximation for $\dmhdt$ is a derived quantity.

Using the optimization techniques detailed in Appendix~\ref{appendix:individual_growth}, we have identified a set of best-fitting parameters for every halo in the samples described in \S\ref{sec:sims}. Figure~\ref{fig:individual_residuals} displays the quality of these fits for both gravity-only N-body simulations and \tng. Each panel shows the logarithmic difference between the simulated and best-fitting MAH plotted as a function of cosmic time; results for samples of halos of different present-day mass are shown in different panels as indicated by the in-panel annotation. The black curves in each panel show the average residual difference, and the shaded band shows the $1\sigma$ scatter; residuals for gravity-only simulations are shown with the solid black curve and purple shaded region; residuals for \tng are shown with the dashed black curve and green shaded region. 

Figure~\ref{fig:individual_residuals} shows that the rolling power-law model defined by Eq.~\ref{eq:individual_mah} gives an unbiased approximation to halo mass assembly with a typical error of 0.1 dex for $t\gtrsim1$ Gyr; residual errors are slightly larger at higher halo mass due to the increased prominence of major mergers. In Appendix~\ref{appendix:subs}, we show that analogous results hold for the case of present-day host halos that have experienced a flyby event at some point in their past history (Figure~\ref{fig:splashback_residuals}), for present-day subhalos (Figure~\ref{fig:subhalo_residuals}), and also for subhalos that have either merged or become disrupted prior to $z=0$ (Figure~\ref{fig:orphan_residuals}). In Figure~\ref{fig:individual_residuals}, we have restricted attention to $t>1$ Gyr for the sake of ensuring good mass resolution across the mass range of all our samples; this cut in cosmic time is driven by our focus on the redshifts $z\lesssim5$ that are most relevant to cosmological surveys, and by the lack of a publicly-available, higher-resolution simulation with homogeneously processed merger trees (see \S\ref{appendix:individual_growth} for further discussion).

\subsubsection{Capturing halo assembly bias}
\label{subsec:assembias}

Even though Figure~\ref{fig:individual_residuals} shows that simulated MAHs are well approximated by the \dmah model, of course the agreement is not perfect, and a question that naturally arises is whether the residual errors are correlated with the cosmic density field. As mentioned in \S\ref{sec:intro}, at fixed present-day mass, the large-scale density of halos exhibits a significant dependence upon assembly history, a phenomenon referred to as {\em halo assembly bias.} This effect is commonly quantified in terms of the two-point cross-correlation between halos and dark matter particles, $\xi_{\rm \delta m}(r).$ For example, for a sample of dark matter halos with the same value of $M_{0}\equiv\mpeak(t_0),$ the quantity $\xi_{\rm \delta m}(r)$ varies with the formation time of the halo, $\tform,$ defined as the first time the halo attains some specified fraction of $M_0.$ The non-zero residual errors shown in Fig.~\ref{fig:individual_residuals} indicate that each halo's approximate and actual values of $\tform$ will differ, and it is plausible that the difference correlates with the density field in a way that alters the $\tform$-dependence of $\xi_{\rm \delta m}(r).$ For example, halos with a merger-rich history tend to reside in denser regions relative to halos with a more quiescent assembly, and the prevalence of mergers may be correlated with the residual errors of our best-fitting approximation.

\begin{figure}
\includegraphics[width=8cm]{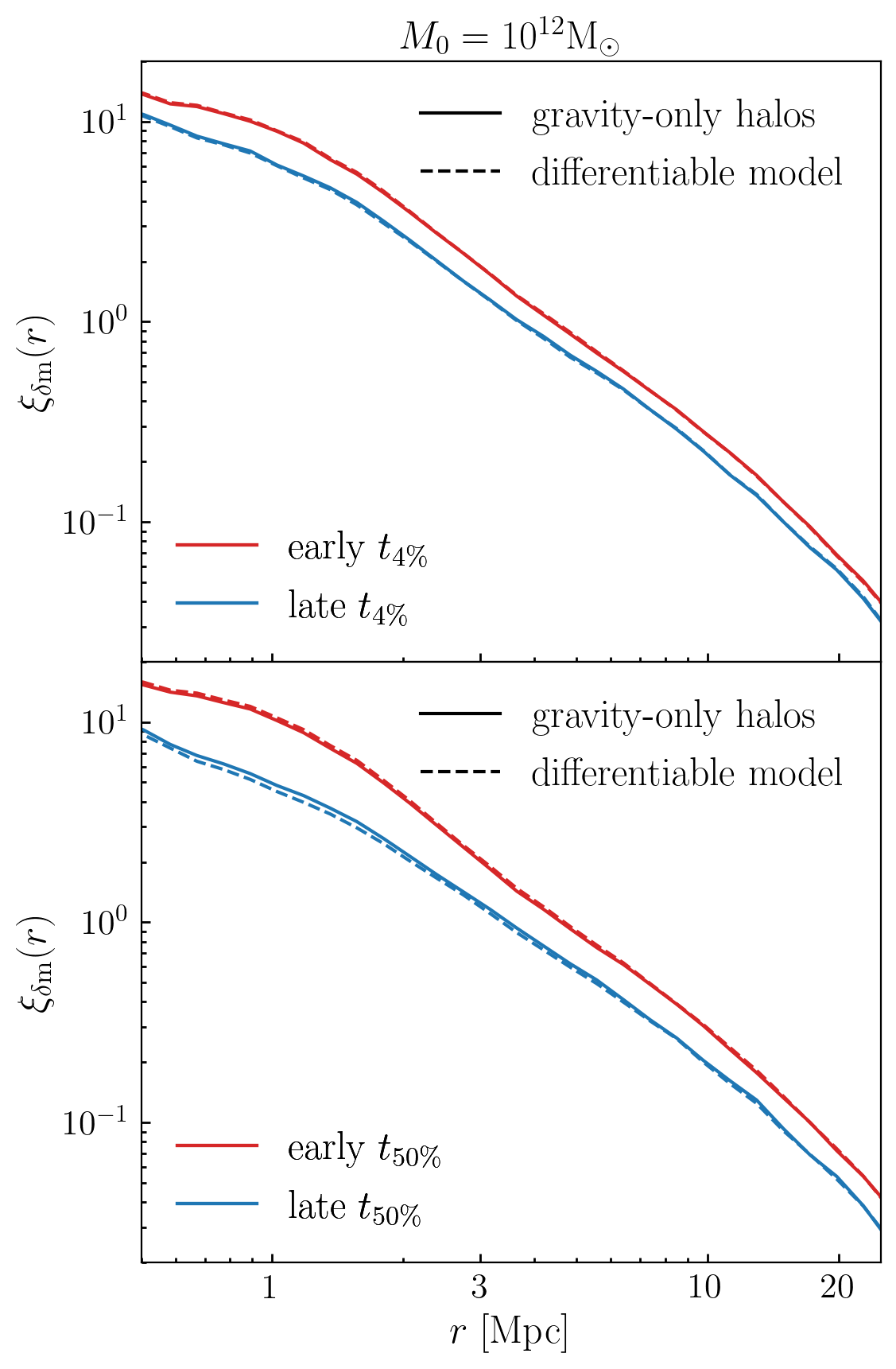}
\caption{{\bf Assembly bias in the \dmah model for individual halo growth.} Using a sample of host halos in the BPL simulation with $\mpeak(t_0)\approx10^{12}\msun,$ we divide the sample in half according to $\tform,$ the median value of halo formation time for the sample, and for each subsample we compute $\xi_{\rm \delta m}(r),$ the cross-correlation between halos and dark matter particles. Red curves show $\xi_{\rm \delta m}(r)$ for early-forming halos, blue curves for late-forming halos; solid curves show results for the case where $\tform$ is computed directly from each halo's simulated merger tree; dashed curves show results for the case where $\tform$ is defined by the differentiable approximation to each halo's assembly history. The top and bottom panels show results for different definitions of $\tform$ as indicated in the legend. The figure demonstrates that the correlation between halo formation time and the density field is retained when simulated merger trees are approximated with our differentiable model.}
\label{fig:assembias}
\end{figure}

To investigate this question, we select a sample of host halos in the BPL simulation with $10^{11.9}\msun\leq\mzero\leq10^{12.1}\msun,$ divide the sample in half according to the median value of $\tform$ for the sample, and use {\tt halotools} \citep{hearin_etal17_halotools} to compute $\xi_{\rm \delta m}(r)$ for each subsample. We have repeated this exercise using the value of $\tform$ computed directly from each halo's simulated merger tree, and compared the clustering to the case when $\tform$ is defined instead based on the differentiable approximation to each individual halo's MAH. We display our results in Figure~\ref{fig:assembias}; the top panel shows $\tform=\tperc{4},$ and the bottom panel shows $\tform=\tperc{50}.$ We find a similar level of agreement for halos of different mass in both gravity-only simulations, and Figure~\ref{fig:assembias_tng} in Appendix~\ref{appendix:tng} shows the analogous result for the case of \tng. We have also verified that we obtain a similar level of agreement between simulation- and model-based clustering when dividing our halo samples based on quartiles of formation time rather than medians.

In summary, the results of \S\ref{sec:individual_halos} and Appendices~\ref{appendix:subs}-\ref{appendix:tng} demonstrate that the 3-parameter fitting function defined by Eq.~\ref{eq:individual_mah} is sufficiently flexible to approximate $\mpeak(t)$ with an accuracy of better than 0.1 dex across nearly all of cosmic time; that this level of accuracy is robust to variations in halo growth due to hydrodynamics and baryonic feedback, and applies equally well to subhalos and splashback centrals; and that the small residual errors of the approximation exhibit an essentially negligible correlation with the large-scale density field.

\section{Modeling the Assembly of Halo Populations with DiffmahPop}
\label{sec:halo_populations}
In the previous section, we presented \dmah, a model for the assembly history of individual dark matter halos defined by Eq.~\ref{eq:individual_mah}. Our model for $\mpeak(t)$ is characterized by three parameters: $\tauc, \aearly,$ and $\alate.$ In this section, we present \dmahpop, a model for $P(\mpeak(t)\vert\mzero),$ the statistical distribution of assembly histories for halos of present-day peak mass, $\mzero.$ Having shown that individual simulated MAHs can be accurately approximated by our parametric fitting function, our approach to modeling $P(\mpeak(t)\vert\mzero)$ is to construct a suitably accurate statistical model for $\parampdf,$ the distribution of our MAH model parameters as a function of $\mzero.$

\begin{figure}
\includegraphics[width=8cm]{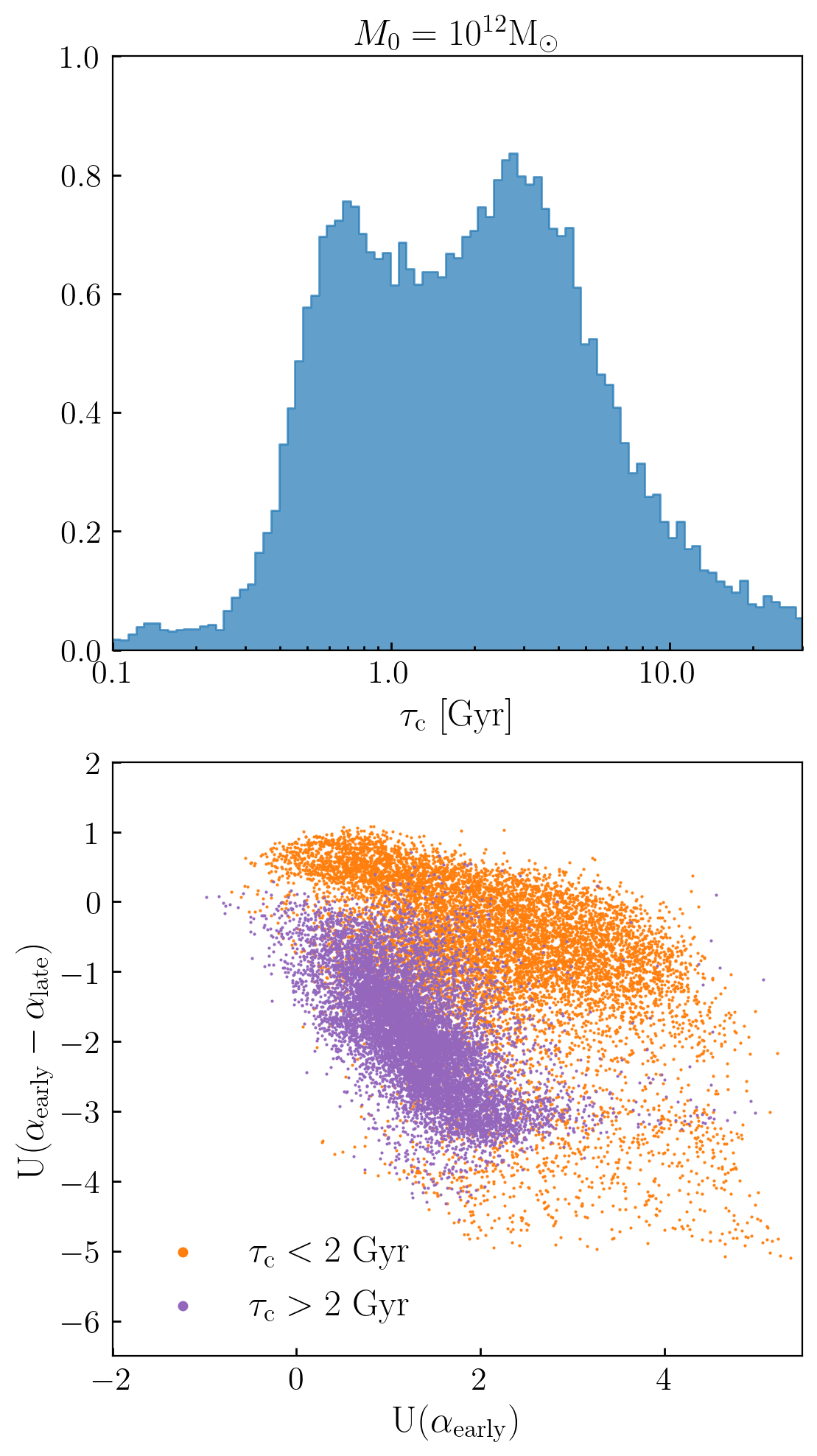}
\caption{{\bf Bimodal distribution of \dmah parameters.} The figure shows the distribution of best-fitting parameters of our model for individual halo growth for a sample of host halos in the BPL simulation with $\mpeak(t_0)\approx10^{12}\msun.$ The top panel shows the distribution of $\tauc,$ which presents a modest bimodality of early-forming halos ($\tauc<2$ Gyr) and late-forming halos ($\tauc>2$ Gyr). The bottom panel shows a scatter plot of $U(\aearly),$ and $U(\aearly-\alate),$ where $U(s)\equiv\ln(\exp(s) - 1),$ with points color-coded according to their value of $\tauc.$ The two panels together highlight the bimodal structure of the three-dimensional probability distribution, $P(\tauc, \aearly, \alate\vert\mzero).$ We leverage this bimodality when building our model for the growth of halo populations.}
\label{fig:mah_pdf_bimodality}
\end{figure}

In order to motivate the form of our model for $\parampdf,$ in Figure~\ref{fig:mah_pdf_bimodality} we show two different cross sections of this multi-dimensional distribution, focusing on BPL halos with $\mzero=10^{12}\msun.$ The top panel shows a histogram of $P(\tauc\vert\mzero),$ which shows a roughly log-normal shape with a modest bimodality in which two distinct sub-populations emerge: earlier-forming halos with $\tauc\lesssim2$ Gyr, and later-forming halos with $\tauc\gtrsim2$ Gyr. In the bottom panel of Figure~\ref{fig:mah_pdf_bimodality}, we show a scatter plot of the two-dimensional distribution of $P(\lge,\lgl\vert\mzero),$ where we define $\lgl=U(\alate),$ the variable $\lge=U(\aearly-\alate),$ and the function $U(s)\equiv\ln(\exp(s) - 1).$ As described in Appendix~\ref{appendix:individual_growth}, the variables $\lge$ and $\lgl$ are the actual quantities that we programmatically vary when seeking best-fitting approximations to the growth of individual halos, as this enforces physical constraints in the approximation to each halo's assembly. In the bottom panel of Figure~\ref{fig:mah_pdf_bimodality}, points with $\tauc>2$ Gyr are shown in purple, and points with $\tauc<2$ Gyr are shown in orange. The color-coding in the bottom panel brings the bimodal structure of $\parampdf$ into sharper relief; while the two-dimensional marginal distribution $P(\aearly, \alate\vert\mzero)$ is rather complex, the same two-dimensional distribution becomes significantly simpler when transforming to the variables $\lge$ and $\lgl,$ and conditioning the distribution upon whether $\tauc<2$ Gyr. In the remainder of this section, we will outline how we have leveraged this apparent bimodality to build a simple two-population model of halo assembly; we refer the reader to Appendix~\ref{appendix:population_growth} for a more expansive discussion of the full three-dimensional distribution, $\parampdf.$

We model $\parampdf$ as a two-population normal distribution of three variables: $\log_{10}\tauc,$ $\lgl=U(\alate),$ and $\lge=U(\aearly-\alate).$ For BPL, MDPL2, and \tng, we find that relative abundance of the two populations is roughly $50\%$, but exhibits a shallow mass-dependence. Each sub-population is described by an independently-defined mean, $\mui,$ and covariance matrix, $\sigi,$ each of which also have a shallow, monotonic mass-dependence, so that $\mui=\mui(\mzero)$ and $\sigi=\sigi(\mzero).$ In calibrating our model parameters for this mass dependence, our principal goal is to faithfully describe $\mahpdf$ and $\dmhdtpdf.$ Thus when optimizing our model parameters, for our target data we use measurements of the first two moments of $\mahpdf$ and $\dmhdtpdf,$ rather than attempting to directly recover the distribution $\parampdf.$  Briefly, we tabulate $\mahpdf$ and $\dmhdtpdf$ using several narrow bins of halo mass, trimming $3\sigma$ outliers before computing mean and variance; we optimize the parameters of our model for $\mui(\mzero)$ and $\sigi(\mzero)$ by minimizing the logarithmic difference between the mean and variance of simulated and predicted values for $\mahpdf$ and $\dmhdtpdf.$ Our differentiable formulation in JAX enables us to optimize our parameters with the gradient-based BFGS algorithm implemented in scipy \citep{broyden_1970_B_in_BFGS,fletcher_1970_F_in_BFGS,goldfarb_1970_G_in_BFGS,shanno_1970_S_in_BFGS}. We refer the reader to Appendix~\ref{appendix:fitting_population_growth} for a detailed description of our application of these techniques.

\begin{figure}
\includegraphics[width=8cm]{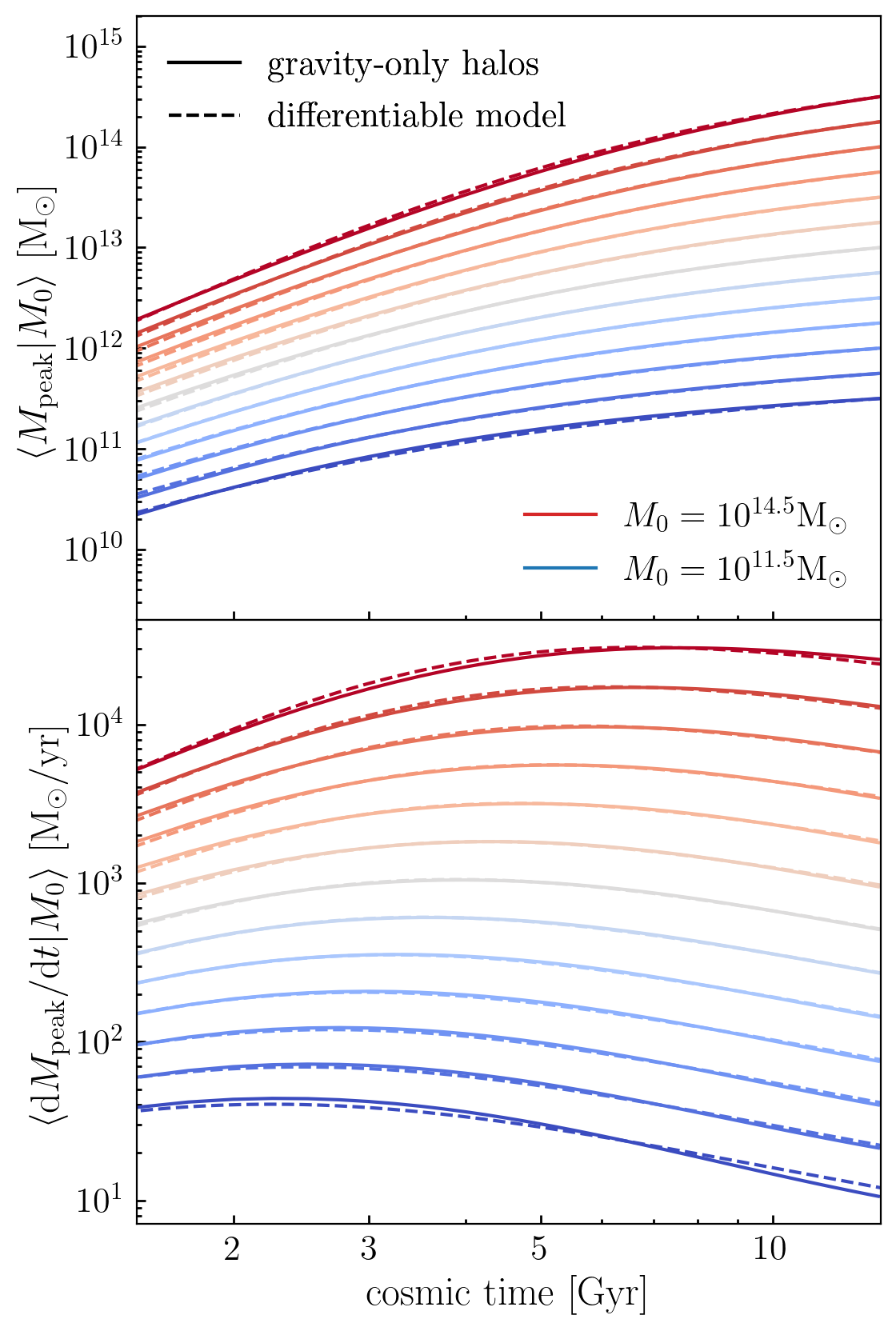}
\caption{{\bf Average assembly history in the \dmahpop model for halo populations.} The top panel shows the average cumulative peak mass as a function of cosmic time, and the bottom panel shows average accretion rates. The histories of halos with different present-day mass are color-coded as indicated in the legend. Solid curves show the average histories of simulated halos; dashed curves show the corresponding quantity approximated by our model for the assembly of halo populations.}
\label{fig:avg_mah_validation}
\end{figure}

\begin{figure*}
\begin{centering}
\includegraphics[width=12cm]{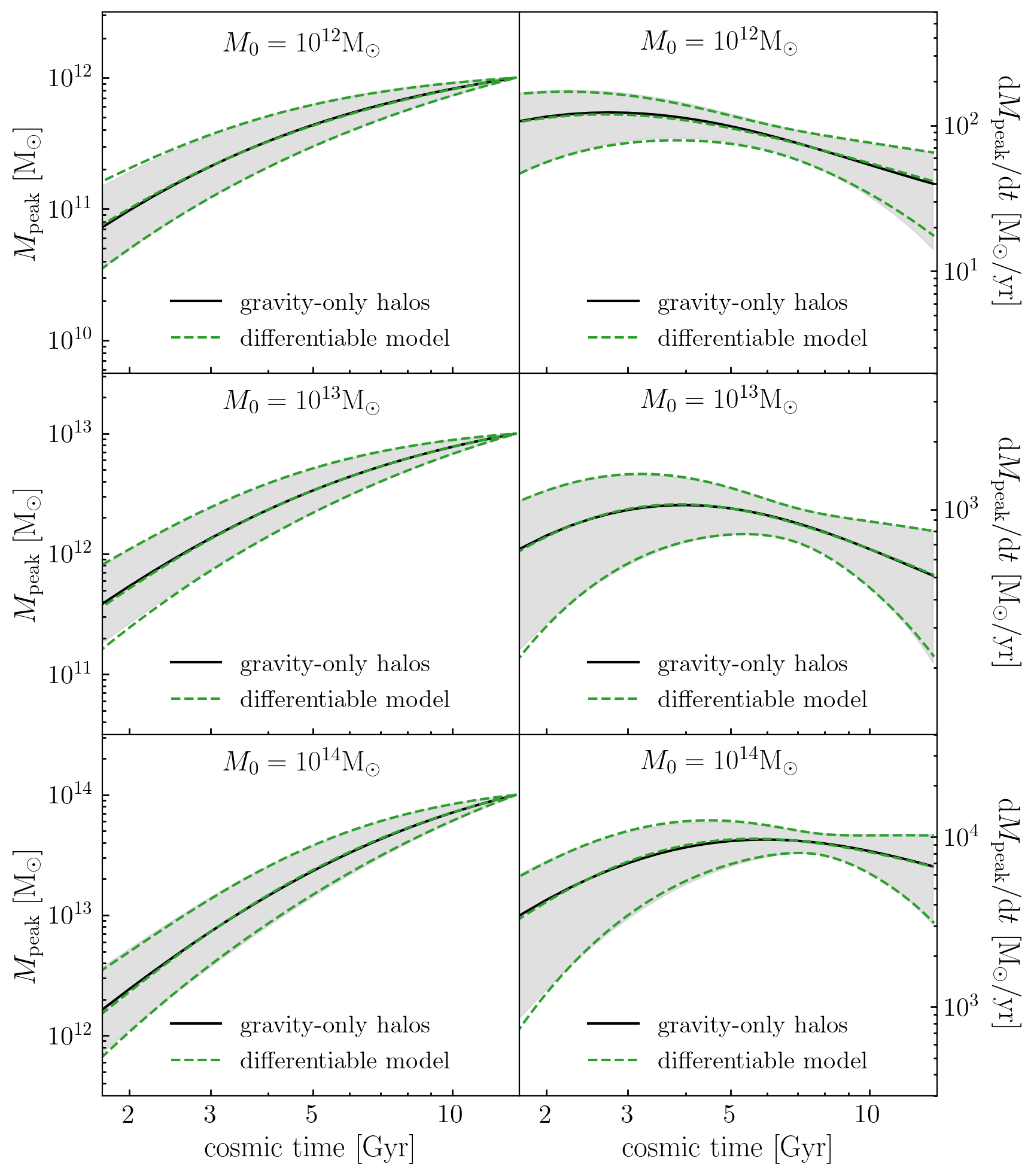}
\caption{{\bf Diversity of assembly histories in the \dmahpop model for halo populations.} In the left column of panels, we plot the history of cumulative mass of simulated halos, $\mpeak(t),$ showing its average value with the solid black curve, and the $1\sigma$ standard deviation with the gray band. The right hand column of panels shows the mean and standard deviation of halo mass accretion rates. The history of halos of different present-day mass are shown in different rows of panels as indicated by the in-panel annotation. In all panels, the dashed green curves show our model for the assembly of halo populations.}
\label{fig:mah_pdf_validation}
\end{centering}
\end{figure*}

\begin{figure}
\includegraphics[width=8cm]{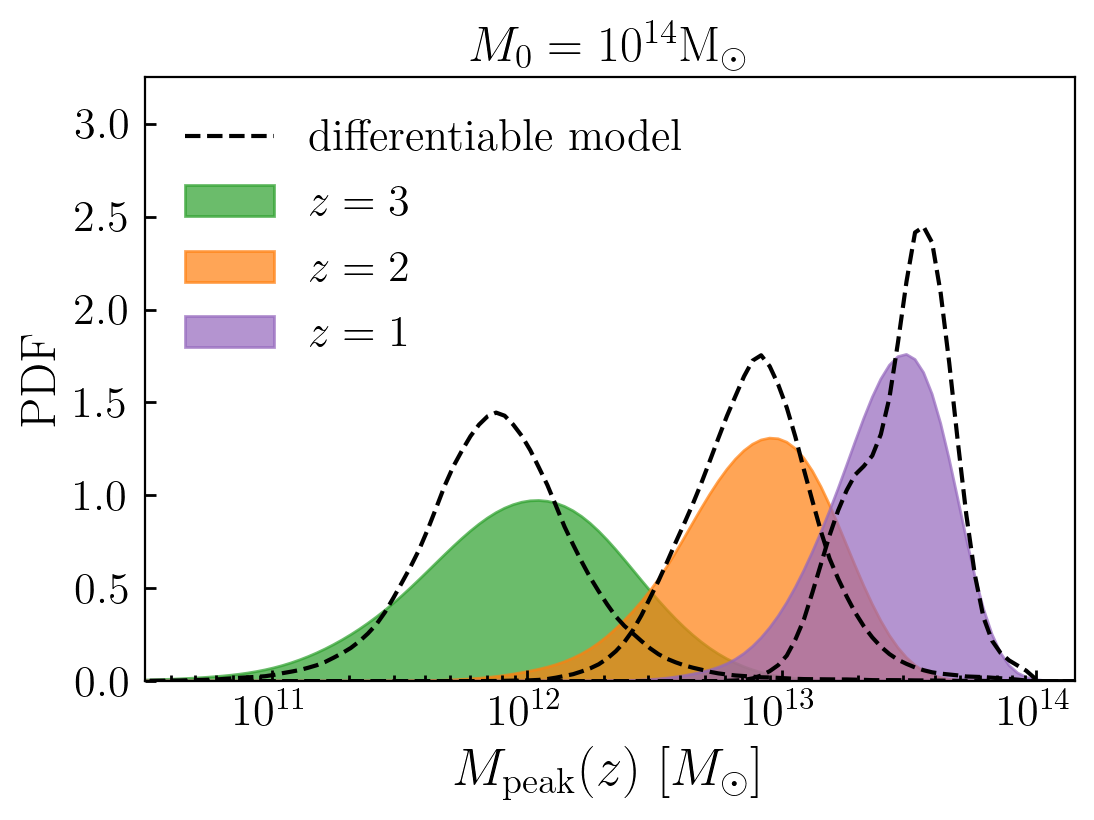}
\includegraphics[width=8cm]{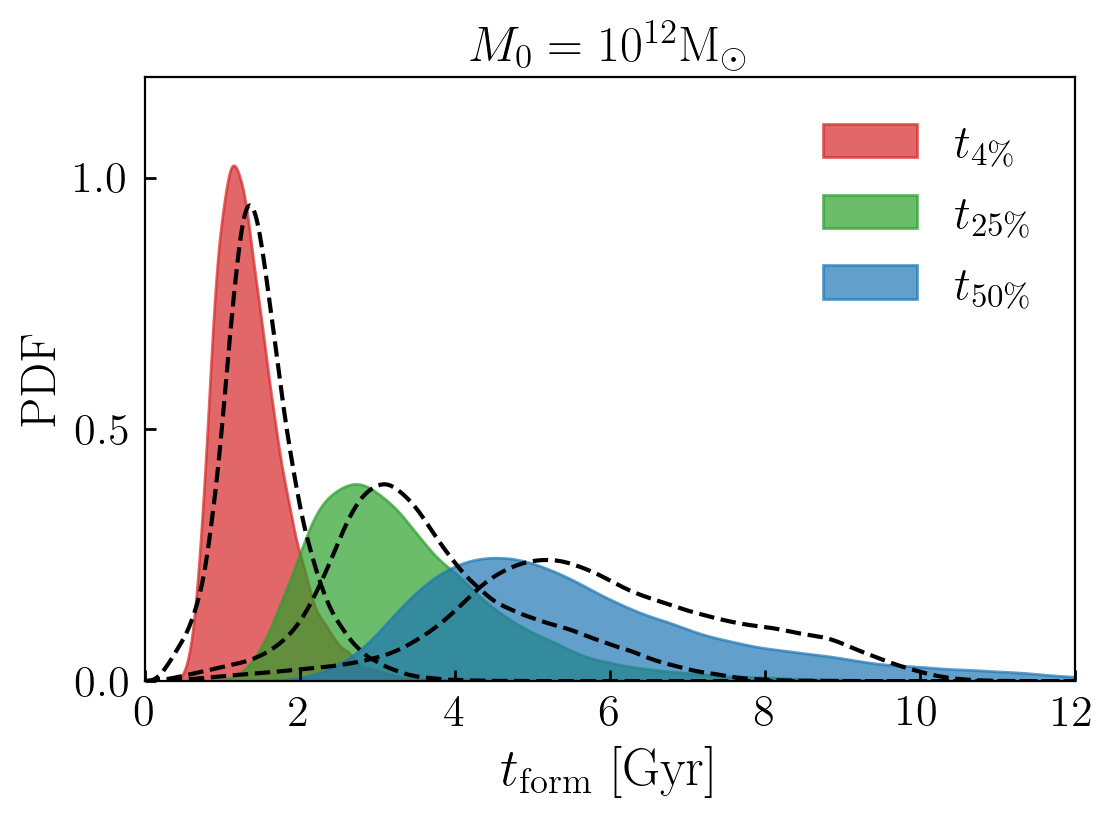}
\caption{{\bf Example comparisons of halo assembly distributions.} In the top panel, we focus on halos with present-day mass $\mzero=10^{14}\msun,$ and show the distribution of masses at higher redshift, $\mpeak(z),$ with results at different redshift color-coded as indicated in the legend. In the bottom panel, for halos with $\mzero=10^{12}\msun,$ we show the distribution of formation times, $\tform,$ using several different definitions as indicated in the legend. Dashed black curves in each panel show the corresponding predictions based on our model for the assembly of halo populations. Plotted quantities shown in this figure are distinct from those shown in Fig.~\ref{fig:mah_pdf_validation} in that here we show comparisons of the full probability distributions of quantities computed directly from simulated merger trees without trimming $3\sigma$ outliers.}
\label{fig:zform}
\end{figure}

In Figure~\ref{fig:avg_mah_validation}, we compare the average assembly of halos in gravity-only simulations to our best-fitting model; the top panel shows the average cumulative peak mass as a function of cosmic time, and the bottom panel shows average accretion rates. Assembly histories for halos of different present-day mass are color-coded as shown in the legend, with dashed curves showing the model, and solid curves showing the corresponding quantity in the simulations; we use BPL for halos with $\mpeak\leq10^{13.5}\msun,$ and MDPL2 for more massive halos.

Several of the basic physical principles of CDM structure formation are on plain display in Figure~\ref{fig:avg_mah_validation}. In the top panels, we can see that for halos of all mass, the slope of $\mpeak(t)$ is steeper at early times relative to late times; the flattening of the slope of $\mpeak(t)$ is a manifestation of the evolution of halos from the fast-accretion regime to the slow-accretion regime. Structure growth in CDM is said to be ``hierarchical'', such that the formation of smaller halos precedes that of more massive halos. The hierarchical nature of halo assembly is visible in both panels, as we can see that the slope of the bluer curves reaches an inflection point at earlier times relative to the redder curves. In the bottom panel we can see the influence of dark energy on halo mass accretion rates: at low redshift when the cosmic acceleration becomes significant, the value of $\dmhdt$ is declining for halos of all mass, with lower-mass halos becoming subject to this trend at earlier epochs.

In Figure~\ref{fig:mah_pdf_validation}, we show that the \dmahpop model for the assembly of halo populations captures both the {\em average} growth of halos, as well as the {\em diversity} of halo assembly. In the left column of panels of Fig.~\ref{fig:mah_pdf_validation}, we plot $\langle\mpeak(t)\vert\mzero\rangle,$ and in the right column of panels we plot $\langle\dmhdt\vert\mzero\rangle;$ in all panels, solid black curves show the quantity taken from the simulation, gray shaded bands show the $1\sigma$ standard deviation of the simulated quantity, and the dashed green curves show the results of our best-fitting model. Each row of panels corresponds to the history of halos with different present-day mass, as indicated by the in-panel annotation.

As mentioned above, the plotted data in Figure~\ref{fig:mah_pdf_validation} are the first and second moments of $\mahpdf,$ which we used directly as target data to calibrate our model parameters. In Figure~\ref{fig:zform}, we show comparisons between simulations and our model predictions for the full distribution, $\mahpdf.$ In the top panel of Fig.~\ref{fig:zform}, we focus on a sample of host halos in MDPL2 with present-day mass of $\mzero=10^{14}\msun,$ and show the distribution of halo masses at several different redshifts as indicated in the legend. In the bottom panel of Fig.~\ref{fig:zform}, we focus on host halos in BPL with present-day mass of $\mzero=10^{12}\msun,$ and show distributions of halo formation time, $\tform,$ defined as the first time the halo attained some specified fraction of its present-day mass; different shaded histograms show different choices for the definition of $\tform$ as indicated in the legend. In both panels, each shaded histogram is accompanied by a dashed black curve showing the predictions of our differentiable model of halo populations. 

Fig.~\ref{fig:zform} does not display the same level of agreement between simulations and model as illustrated in Figure~\ref{fig:mah_pdf_validation}, because in Fig.~\ref{fig:zform} we show the full probability distributions taken directly from the simulated merger trees without trimming $3\sigma$ outliers. We refer the reader to \S\ref{appendix:population_optimization} for a detailed description of our outlier exclusion algorithm, which has the effect of excluding 2-3\% of halos at each mass that have recently experienced an unusually large major merger. Thus the shaded histograms in Fig.~\ref{fig:zform} include contributions from short-term fluctuations associated with recent mergers, whereas we optimized our model for halo populations using a sample that excludes such contributions. This difference can be seen in two different manifestations in Fig.~\ref{fig:zform}. First, the PDFs shown in the shaded histograms have slightly broader support relative to our model, reflecting the increased variance associated with the inclusion of outliers. Second, the shaded histograms tend to be shifted towards slightly earlier redshifts relative to the model; this second effect is due to our election to model the evolution of {\em peak} halo mass, $\mpeak(t),$ whose associated formation times are asymmetrically shifted to earlier epochs by major mergers. Both of these effects are visible in Fig.~\ref{fig:zform}. The full distributions for Milky Way mass halos shown in the bottom panel display a somewhat better level of agreement relative to the full distributions for cluster-mass halos shown in the top panel; this is sensible since the assembly of more massive halos tends to be more rich in mergers.


\section{Discussion \& Future Work}
\label{sec:discussion}
We have presented a new model for the assembly history of individual dark matter halos, \dmah. In our model, halo mass is a power-law function of cosmic time with a rolling index, $\mpeak(t)\propto t^{\alpha(t)},$ where $\alpha(t)$ is a sigmoid function that smoothly transitions halo growth from the early-time, fast-accretion regime, to the late-time, slow-accretion regime (see Eqs.~\ref{eq:individual_mah}-\ref{eq:sigmoid}). We note that we have chosen to model {\em peak} halo mass, $\mpeak(t),$ the largest mass attained by the halo up until the time $t,$ and so our model is not intended to capture the instantaneous halo mass, $\mhalo(t),$ which is subject to additional physical effects whose treatment is beyond the scope of the present paper.

The results in \S\ref{sec:individual_halos} in the main body of the paper demonstrate that for host halos identified with Rockstar over a wide range of present-day mass, $\mzero\equiv\mpeak(z=0)\gtrsim10^{11}\msun,$ the \dmah model faithfully describes the mass assembly of individual halos over nearly all of cosmic time, $t\gtrsim1$ Gyr, with a typical accuracy of $\sim0.1$ dex or better. In Appendix~\ref{appendix:subs}, we show that our model for $\mpeak(t)$ enjoys the same level of accuracy regardless of whether the host halo experiences a splashback event in its past history, or whether the halo eventually becomes a subhalo of a more massive host, including cases where the subhalo is disrupted or merges with its host. In Appendix~\ref{appendix:tng} (together with Fig.~\ref{fig:individual_residuals}), we show that our model achieves the same level of accuracy in its description of $\mpeak(t)$ for both host halos and subhalos in \tng, a full-physics hydrodynamical simulation whose halos and merger trees were identified with entirely different algorithms from those used in the analysis of the gravity-only simulations we consider.

We have additionally presented a new model for the assembly of halo populations, \dmahpop. In typical empirical approaches to modeling halo assembly, average halo growth is directly parameterized with a fitting function, and variance in halo assembly is modeled simply as random scatter. Our alternative approach is formulated by using the \dmah model for individual halo growth as its fundamental basis, and by building a model for the statistical distribution of the \dmah parameters that regulate individual halo assembly. As a result, \dmahpop not only reproduces $\langle\mpeak(t)\vert\mzero\rangle,$ but also a cosmologically realistic {\em diversity} of individual halo trajectories across time, $\mahpdf.$ The \dmahpop model for the growth of halo populations is differentiable courtesy of its implementation in the autodiff library JAX, and accurately captures the average mass accretion rate across cosmic time, $\langle\dmhdtt\vert\mzero\rangle,$ as well as the diversity of accretion rates $\dmhdtpdf.$ As a convenience, our publicly available python code includes functionality to generate a Monte Carlo realization of the diversity of halos with the same $\mzero.$

The accuracy of our approximation may not be surprising as previous results have shown that halo mass is predominantly assembled via a combination of smooth accretion and mergers with a very small mass ratio \citep[e.g.,][]{genel_etal10}. However, our model could naturally be used as the basis of empirical techniques to predict merger rates and/or the subhalo mass function at the time of accretion (also referred to as the {\em unevolved subhalo mass function}, or USHMF hereafter). For example, in \citet{yang_mo_zhang_vdb_2011}, the authors developed an analytical model that accurately predicts the USHMF; as the basis of their model, they used the fitting function of \citet{zhao_etal03} for the {\em median} assembly history of dark matter halos, supplementing this with a simple log-normal distribution to describe random scatter in the main-branch mass as a function of time. The techniques in the present work could be used to improve upon such an approach, as our model provides an accurate description of $\mahpdf,$ as shown in Figure~\ref{fig:mah_pdf_validation}. Such an extension could also prove fruitful when used in concert with subhalo orbital evolution codes \citep[e.g.,][]{zentner_etal05,jiang_etal2021_satgen1}. Recent progress in modeling subhalo orbital evolution has demonstrated the importance of accounting for the evolution of the host halo potential \citep{ogiya_etal21}, and so deploying our model in this context could be leveraged to capture physically realistic accretion-rate correlations in the halo-to-halo variance of substructure abundance and orbits across cosmic time \citep[see, e.g.,][]{jiang_vdb_2017}. Along similar lines, it has recently been shown that scatter in the thermal Sunyaev-Zel'dovich effect is largely driven by variance in the assembly histories of cluster-mass halos \citep{green_etal20}, and so our model may also help improve the predictive power of models for the mass-observable relation of galaxy clusters.

While previous work has demonstrated that the cosmology-dependence of halo assembly is relatively simple \citep[e.g.,][]{zhao_jing_mo_2009}, all results in the present work are limited to a fixed set of cosmological parameters that closely match those in \citet{planck14b}. In \citet{vdb_etal14}, it was shown that halo assembly histories have a universal form controlled entirely by the cosmology-dependence of cosmic time, and so it may be relatively straightforward to extend our model for $\mahpdf$ to capture the influence of cosmological parameters on halo growth. Such an extension could be directly incorporated into frameworks that model cosmology-dependence via physically-motivated rescalings \citep[e.g.,][]{angulo_white_2010,renneby_etal18,arico_angulo_etal20}. In principle, our model could also be used in applications with emulators built upon cosmological suites of simulations \citep[e.g.,][]{heitmann_etal16_mira_titan,derose_etal19_aemulus1,nishimichi_etal19_dark_quest}, although this would require suites with high mass-resolution that include merger trees, which remains uncommon \citep[although see][for recent progress]{contreras_etal20}.

Our restriction to cosmological parameters similar to \citet{planck14b} is just one example of how in the present work, we have not attempted to provide a comprehensive study of $\mahpdf.$ As another example, even though we have provided separate calibrations of our model for halos in gravity-only simulations and in \tng, we have not attempted a detailed characterization of the differences between the assembly histories of halos in the two simulations. For such a purpose, a simulation suite such as CAMELS \citep{villaescusa_navarro_etal20_camels1} that spans a range of baryonic feedback parameters would enable a more rigorous study than an uncontrolled comparison between \tng and N-body halos.  Furthermore, when calibrating our models for $\mahpdf,$ we have focused on the mass range $10^{11.5}\msun\leq\mpeak\leq10^{14.5}\msun$ as this range of halo masses contains thousands of halos that are resolved by over $2000$ particles in the N-body simulations we use; however, a series of nested-resolution boxes would be considerably more effective towards the goal of extending the mass range and conducting rigorous resolution tests. As discussed in \S\ref{appendix:individual_growth}, we have similarly focused on the range of cosmic time $t>1$Gyr to avoid potential resolution issues in our simulations, but our work could be extended to redshifts $z>5$ with a simulation suite that would permit such resolution tests. Improvements such as these would require a dedicated effort along the lines of what has been done to calibrate models of the halo mass function \citep[e.g.,][]{jenkins_etal01_mass_function,tinker_etal08_mass_function,mcclintock_etal19_mass_function,bocquet_etal20_mass_function}; we thus consider these improvements beyond our present scope, as here we merely demonstrate that a modeling approach such as ours has capability to deliver a precision tool.

When fitting the growth history of large samples of simulated halos, we find that the distribution of best-fitting parameters exhibits a clear bimodality, with earlier-forming and later-forming halos occupying distinct regions of the three-dimensional parameter space of $\aearly, \alate,$ and $\tauc.$ Bimodal halo growth has been reported previously in \citet{shi_wang_mo_etal2018} in terms of the distribution of halo formation times, although this previous finding applied only to {\em infall halos} that eventually became subhalos of some more massive host. The bimodality we find pertains to {\em host} halos at $z=0,$ applies to halos of all mass, persists even when excluding halos that have previously experienced a flyby event, and is present with comparable strength in both gravity-only simulations and \tng. We have leveraged this bimodal structure in building our analytical model for $\mahpdf,$ as capturing this feature considerably simplifies the task of achieving the level of agreement with simulations shown in Fig.~\ref{fig:mah_pdf_validation}, particularly for the case of the second moment.

The two-point clustering of dark matter halos of the same mass has well-known dependence upon halo formation time, $\tform,$ a phenomenon known as {\em halo assembly bias}. In Figure~\ref{fig:assembias}, we showed that the $\tform$-dependence of halo clustering remains the same whether one uses the actual formation time taken directly from simulated merger trees, or the value of $\tform$ taken from our best-fitting differentiable model. This result implies that our model can approximate halo growth in such a way that residual errors in the fit are uncorrelated with the density field on large scales. This finding is especially interesting in light of the relationship between assembly bias and halo concentration \citep{wechsler_etal2006}, the latter of which is closely connected to assembly history \citep{wechsler_etal2002,chen_etal20}, and is known to be strongly influenced by mergers \citep[e.g.,][]{wang_zentner_etal_2020}. We will investigate this issue systematically in follow-up work to the present paper.

As discussed in \S\ref{sec:intro}, halo assembly history plays a fundamental role in contemporary semi-analytic models of galaxy formation (SAMs), since the star formation rate of a main sequence galaxy is commonly assumed to be proportional to the mass accretion rate of its parent halo. We have formulated our model for the assembly of halo populations with numerous applications to SAMs squarely in mind. For example, our model has potential to considerably accelerate efforts to explore the portions of SAM parameter space that regulate $\epsilon_{\rm eff},$ the efficiency with which galaxies transform accreted mass into stars, since running SAMs on main-branch histories alleviates the need to solve galaxy formation ODEs along every branch of the merger tree. On the one hand, this computational speedup comes at the cost of neglecting the role of mergers on SAM-predicted galaxy properties. On the other hand, the capability of our model to generate a physically realistic diversity of smooth, individual halo assembly histories enables a rather comprehensive investigation of the specific role of mergers in shaping the properties of the galaxy population. We intend to pursue these directions in future work based on the Galacticus semi-analytic model \citep{benson_galacticus_2012}; Galacticus already includes an independent implementation of our model in its publicly available source code\footnote{\url{https://github.com/galacticusorg/galacticus/wiki}}, which may be useful in its own right for researchers more accustomed to programming in Fortran rather than Python.

As the principal aim of our future work, we are using our model for halo assembly as the foundation of a new approach to forward modeling cosmological structure formation, Surrogate Modeling the Baryonic Universe (SBU). The basis of SBU is a minimally flexible parameterized family of solutions to the differential equations of galaxy formation; we make predictions for galaxy populations by building a statistical model for a cosmologically realistic diversity of individual galaxy histories \citep{alarcon_etal21}. The approach taken in the present work to modeling halo populations essentially provides a template for this framework.

This forward modeling effort is already well underway. In \citet{chaves_montero_hearin_2020_sbu1}, we studied the influence of star formation history (SFH) on the broadband photometry of galaxies and found that, despite the complexity of the processes that impact optical color, physically-motivated variations in SFH correspond to changes in a unique direction in color-color space. In a follow-up paper \citep{chaves_montero_hearin_2021_sbu2}, we studied how the presence of unresolved SFH fluctuations impact model predictions for broad-band photometry, finding that the statistical distributions of the broad-band colors of a galaxy population are quite insensitive to short-term variability. These results taken together indicate that the influence of {\em halo} assembly upon galaxy photometry is likely to be simple, and that the absence of an explicit treatment of halo mergers in our model may not pose a problem for the ability of SBU models to make accurate predictions for the color distributions observed by imaging surveys such as the Dark Energy Survey\footnote{\url{https://www.darkenergysurvey.org/}} \citep[DES,][]{des} and the Legacy Survey of Space and Time\footnote{\url{http://www.lsst.org/}} \citep[LSST,][]{ivezic_etal08, lsst_science_book}. Furthermore, the results shown in Figure~\ref{fig:assembias} indicate that SBU predictions for color-dependent clustering are also likely to be unbiased by the smooth nature of our model for halo assembly.

We expect that the differentiable formulation of our halo assembly model based on the JAX autodiff library will play a critical role in this forward-modeling program. Modern analyses of large-scale structure rely upon a considerable number of ``nuisance parameters'' in order to ensure that cosmological inference is robust to systematic uncertainty. Conventional Bayesian techniques such as MCMC scale poorly with the dimension of the model parameter space, and since the statistical precision of cosmological surveys will dramatically improve in the 2020s, the parameter space of models used to analyze large-scale structure measurements will only continue to increase. The capability to make differentiable predictions addresses this growing problem, since gradient-based optimization techniques such as Adam are extremely efficient even in very high dimensions \citep{kingma_ba_adam_2015}, as are gradient-based inference methods such as Hamiltonian Monte Carlo \citep{hoffman_gelman_2014_nuts}. In the present work, we have shown that not only is our model for individual halo growth differentiable, but using the weighted sampling techniques described in Appendix~\ref{appendix:fitting_population_growth}, so are one-point function predictions based on our model for the assembly of halo populations. In a closely related paper to the present work \citep{hearin_etal21_shamnet}, we will introduce a new theoretical framework for making differentiable predictions for two-point functions such as galaxy clustering and lensing, using abundance matching \citep{kravtsov_etal04} as a toy model for the galaxy-halo connection. The halo assembly model presented here will provide a core ingredient to our program to make differentiable, multi-wavelength predictions for large-scale structure that are physically self-consistent across cosmic time.

\section{Summary}
\label{sec:summary}

We conclude by summarizing our primary results:
\begin{enumerate}
    \item We have introduced \dmah, a new fitting function for $\mpeak(t),$ the evolution of cumulative peak mass of individual dark matter halos; our model approximates halo assembly as a power-law function of cosmic time with rolling index, $\mpeak(t)\propto t^{\alpha(t)}$ (see Eq.~\ref{eq:individual_mah}). Using both gravity-only simulations and \tng, we have demonstrated that the \dmah model can approximate halo growth with an accuracy of 0.1 dex over the range $t\gtrsim1\,{\rm Gyr}$ for halos of present-day mass $\mzero\gtrsim10^{11}\msun.$
    \item We find that the connection between halo assembly and the large-scale density field, known as halo assembly bias, is entirely captured by \dmah, and that residual errors of our differentiable approximation to halo growth exhibit a negligible correlation with the density field on large scales.
    \item We have introduced a new model for the growth of halo populations, \dmahpop, and shown that our model faithfully reproduces the evolution of average halo mass, $\langle\mpeak(t)\vert\mzero\rangle,$ and average mass accretion rate, $\langle\dmhdtt\vert\mzero\rangle,$ for cosmic times $t\gtrsim1\,{\rm Gyr}.$ The \dmahpop model additionally captures the {\em diversity} of halo mass assembly, $\mahpdf,$ as well as the diversity of accretion rates, $\dmhdtpdf.$
    \item Our python code, {\tt diffmah}, is publicly available and can be installed with pip or conda. The repository for our source code is on GitHub, \url{https://github.com/ArgonneCPAC/diffmah}, and includes Jupyter notebooks providing demonstrations of typical use cases. A parallelized script in the {\tt diffmah} repository can be used to fit the assembly histories of individual simulated halos with the \dmah parameters. The {\tt diffmah} code provides a differentiable description of $\mahpdf$ and $\dmhdtpdf$ for both gravity-only simulations and \tng, and also includes a convenience function that can be used to generate Monte Carlo realizations of cosmologically realistic populations of halo assembly histories. Precomputed fits for hundreds of thousands of halos in the BPL, MDPL2, and \tng simulations are available at \url{https://portal.nersc.gov/project/hacc/aphearin/diffmah_data/}.
\end{enumerate}

\section*{Acknowledgements}
This manuscript has been revised since the version originally published by the Open Journal of Astrophysics (OJA). In this revised version, a change to the nomenclature reflects an explicit distinction between the Diffmah model of individual halo growth, and the DiffmahPop model of the growth of halo populations.

We thank Raul Angulo, Andrew Benson, Joe DeRose, Alexie Leauthaud, and Daisuke Nagai for useful discussions. Thanks also to Michael Buehlmann and Aurora Cossairt for feedback on an early release of our source code, and Benedikt Diemer for kindly providing downsampled particles for \tng. APH thanks Lauren Mastro for keeping the WWOZ Sunday Gospel Show going throughout the pandemic.

We thank the developers of {\tt NumPy} \citep{numpy_ndarray}, {\tt SciPy} \citep{scipy}, Jupyter \citep{jupyter}, IPython \citep{ipython}, scikit-learn \citep{scikit_learn}, JAX \citep{jax2018github}, and Matplotlib \citep{matplotlib} for their extremely useful free software. While writing this paper we made extensive use of the Astrophysics Data Service (ADS) and {\tt arXiv} preprint repository. The authors gratefully acknowledge the Gauss Centre for Supercomputing e.V. (www.gauss-centre.eu) and the Partnership for Advanced Supercomputing in Europe (PRACE, www.prace-ri.eu) for funding the MultiDark simulation project by providing computing time on the GCS Supercomputer SuperMUC at Leibniz Supercomputing Centre (LRZ, www.lrz.de). The Bolshoi simulations have been performed within the Bolshoi project of the University of California High-Performance AstroComputing Center (UC-HiPACC) and were run at the NASA Ames Research Center.

Work done at Argonne National Laboratory was supported under the DOE contract DE-AC02-06CH11357. We gratefully acknowledge use of the Bebop cluster in the Laboratory Computing Resource Center at Argonne National Laboratory.

\bibliographystyle{aasjournal}
\bibliography{bibliography}

\appendix
\renewcommand{\thefigure}{A\arabic{figure}}
\section{Fitting the Assembly History of Individual Halos with Diffmah}
\label{appendix:individual_growth}

In this appendix, we give a detailed description of \dmah, our parametric model for the mass assembly of individual halos. As outlined in \S\ref{sec:individual_halos}, the \dmah model for individual halo growth is based on Eq.~\ref{eq:individual_mah}, which describes a power-law scaling between peak halo mass and cosmic time, $\mpeak=\mzero(t/t_0)^{\alpha(t)},$ where the power-law index $\alpha(t)$ smoothly rolls between asymptotic values $\aearly$ and $\alate$ according to the sigmoid function in Eq.~\ref{eq:sigmoid}. Thus there are a total of six numbers that fully characterize our model for individual halo growth: the normalization, $\mzero,$ the present-day cosmic time, $t_0,$ and the four sigmoid parameters, $\aearly, \alate, k,$ and $\tauc.$ For all halos in each simulation, we hold $t_0$ fixed to the present-day age of the universe in the simulation, and $\mzero$ fixed to the value $\mpeak(t_0)$ in the simulation. All results in the paper also hold $k$ fixed to 3.5; to determine this particular value, we ran the optimization algorithm described below while allowing $k$ to be a free parameter, and then observed no appreciable changes in the quality of the fits when holding $k$ fixed to any value in its typical best-fitting range, $2\lesssim k\lesssim 5.$ Thus the \dmah model for individual halo growth has a total of 3 free parameters: $\aearly, \alate,$ and $\tauc.$

When evaluating the sigmoid behavior of the power-law index, we implement this scaling relation in logarithmic time, $x\equiv\log_{10}t,$ rather than linear time. Thus in practice, we calculate the power-law index $\alpha(t)$ via:
\beq
\label{eq:powerlaw_implementation}
\alpha(t)=\alpha(10^x)=\aearly + \frac{\alate-\aearly}{1 + \exp(-k(x-x_0))}.
\eeq

In order to identify an optimal set of parameters for each halo, we searched our three-dimensional model parameter space, ${\theta},$ for the combination of $\aearly, \alate, \tauc$ that minimizes the quantity $\mathcal{L}_{\rm MSE},$ defined as
\beq
\label{eq:mse_loss}
\mathcal{L}_{\rm MSE}\equiv\frac{1}{N}\sum_{i}\left(w(t_i; {\theta})-v(t_i)\right)^2,
\eeq
where $w$ and $v$ are the base-10 logarithm of the predicted and simulated values of $\mpeak$ evaluated at a set of $N$ control points, $t_i.$ For the control points of each halo, we use the collection of snapshots in the halo's merger tree after a time $t_{\rm min},$ defined as
\beq
\label{eq:tminfit}
t_{\rm min}\equiv{\rm max}(t_{\rm cut},\ t_{\rm thresh}, t_{\rm\delta m}),
\eeq
where we set $t_{\rm cut}=1$ Gyr, and the quantity $t_{\rm thresh}$ is the most-recent simulated snapshot where the halo mass falls below a simulation-dependent threshold mass, $M_{\rm thresh};$ for the MDPL2 simulation, we use $M_{\rm thresh}=10^{11.25}\msun,$ and for the Bolshoi-Planck and TNG simulations we use $M_{\rm thresh}=10^{10}\msun.$

\begin{figure}
\includegraphics[width=8cm]{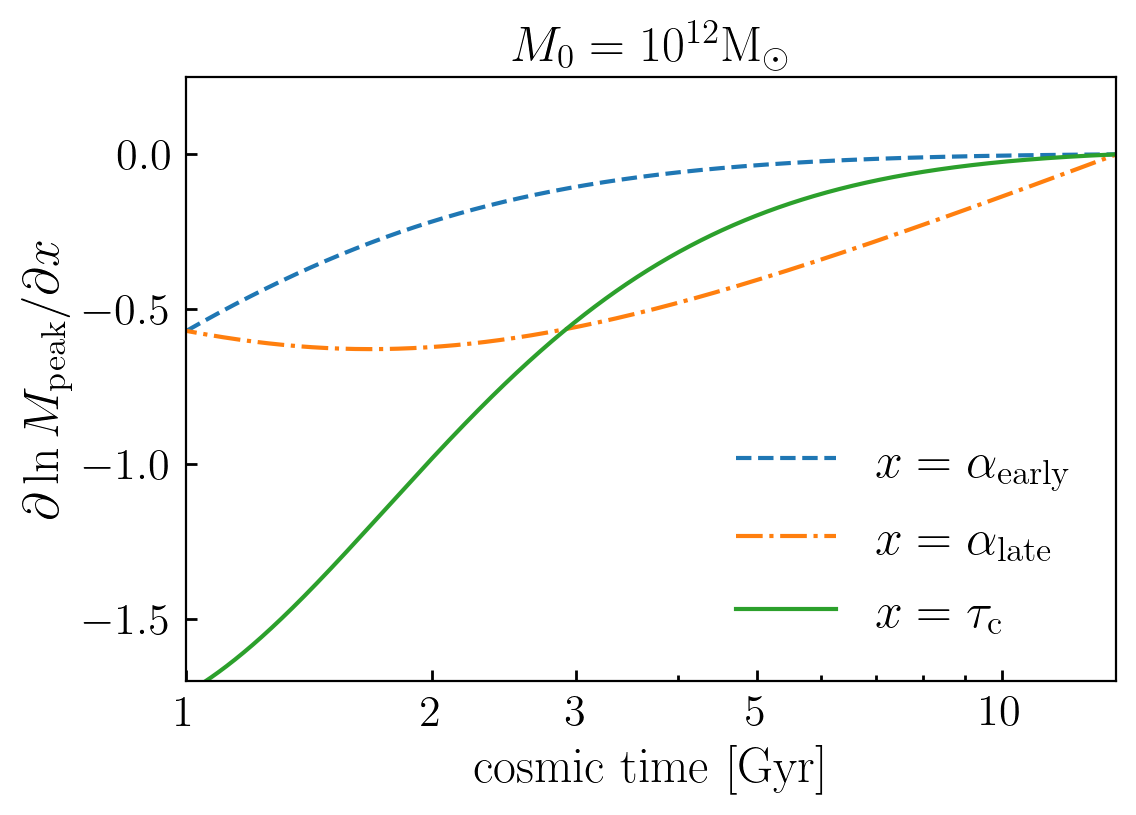}
\caption{{\bf Gradients of the \dmah parameters.} The figure shows logarithmic derivatives of $\mpeak(t)$ with respect to the three parameters of our model for individual halo growth. This figure is similar to Fig.~\ref{fig:diffmah_parameter_variations}, only here we show the behavior of the autodiff-computed gradients that we use to determine the best-fitting parameters describing the assembly of each simulated halo.}
\label{fig:diffmah_parameter_gradients}
\end{figure}

In Eq.~\ref{eq:tminfit}, the quantity $t_{\rm\delta m}$ refers to the most-recent snapshot where the halo mass falls below some fraction $f_{\rm\delta M}$ of its present-day mass; for all simulations, we set $f_{\rm\delta M}=10^{2.5}.$ In each of the three simulations we studied, for a surprisingly large number of halos, we find that at early times, $t\lesssim3$Gyr, the behavior of $\mpeak(t)$ suddenly drops like a stone by an order of magnitude or more from one snapshot to the next, even for halos that should nominally be resolved by hundreds of particles. We found through experimentation that the inclusion of the term $t_{\rm\delta m}$ in Eq.~\ref{eq:tminfit} is reasonably effective at limiting the influence of this phenomenon on the best-fitting parameters describing the halo MAH. We suspect that this behavior is a reflection of the difficulty of halo-finding when the high-redshift density field resembles a sea of shallow peaks, however, the root cause of this phenomenon is unclear, and other origins of this phenomenon are certainly plausible. For example, recent work on structure formation has uncovered surprisingly stringent resolution demands of N-body simulations for science applications that require robust tracking of the evolution of halo substructure \citep{vdb_etal18a, vdb_etal18b, ogiya_etal19}. It could be the case that analogous results apply to the simulation requirements associated with reliable tracking of halo mass accretion history in the first Gyr of cosmic time; alternatively, halo assembly at high redshift may simply require an even more flexible fitting function than the smoothly rolling power-law model adopted here. As discussed further in \S\ref{sec:discussion}, the most reliable way to address this and related issues is through a dedicated resolution-requirement study that uses homogeneously-processed, nested simulation boxes spanning a wide range of resolution. Since such a simulation suite is not currently publicly available, and since our own science applications are focused on the redshift range $z\lesssim5$ that is most relevant to present-day and near-future cosmological surveys, we have opted to relegate further study of this issue to future work. 

To minimize $\mathcal{L}_{\rm MSE}$ for each halo, we use the JAX implementation of the Adam algorithm \citep{kingma_ba_adam_2015}, which is a gradient descent technique with an adaptive learning rate. Our use of Adam requires calculating the gradients $\partial\mathcal{L}_{\rm MSE}/\partial\theta,$ which is facilitated by our JAX implementation, allowing us to efficiently compute these gradients to machine precision without reliance upon finite differencing methods. We show example gradients of our model predictions for $\mpeak(t)$ in Figure~\ref{fig:diffmah_parameter_gradients}. To tune the performance of the optimization beyond the default settings recommended in \citet{kingma_ba_adam_2015}, we use 2 successive burn-in cycles with a relatively large step-size parameter of 0.25 for $\sim50$ updates, followed by a final cycle for an additional $\sim200$ updates using a step-size parameter of 0.01.

\subsection{Imposing physical constraints on individual halo growth}
\label{subsec:constraints}

\begin{figure}
\includegraphics[width=8cm]{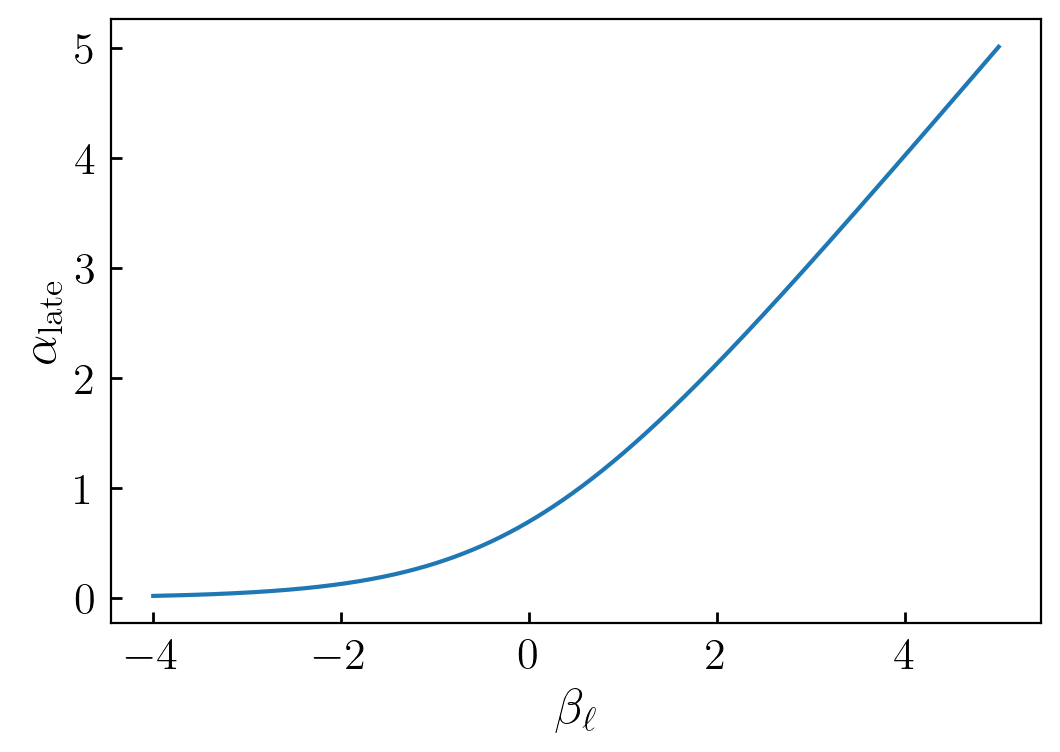}
\caption{{\bf Illustration of physical constraints on $\alate.$} The figure shows the behavior of the softplus function that controls the relationship between $\alate$ and $\lgl,$ the actual variable that we vary in the gradient descent algorithm used to fit the assembly history of each individual halo. We use a similar technique to ensure $0<\alate<\aearly,$ and a simple logarithmic transformation to enforce $\tauc>0.$ These transformations guarantee that the growth history of each model halo will always increase monotonically and transition from a fast-accretion phase to a slow-accretion phase, even for outlier halos with unusual merger trees. See Eq.~\ref{eq:softplus_constraints} and associated discussion for details.}
\label{fig:softplus}
\end{figure}

When minimizing $\mathcal{L}_{\rm MSE}$, the parameters we actually vary in the gradient descent are $\lge,\, \lgl,$ and $x_0,$ where the relationships between these parameters and the quantities that directly appear in Eq.~\ref{eq:individual_mah}, $\aearly,\, \alate,$ and $\tauc,$ are defined as follows:
\beq
\label{eq:softplus_constraints}
x_0 & = & \log_{10}\tauc\nonumber\\
\alate & = & \mathcal{S}(\lgl)\\
\aearly & = & \alate + \mathcal{S}(\lge)\nonumber\\
\mathcal{S}(z) &\equiv& \ln(1 + \exp(z))\nonumber
\eeq
The function $\mathcal{S}(z)$ is the softplus function, which is strictly positive across the real line,  exhibits approximately linear behavior for $s\gtrsim1,$ and asymptotically approaches zero as $s\rightarrow-\infty.$ For any positively-valued $z,$ the inverse of the softplus function is well-defined and computed as $\mathcal{S}^{-1}(z)\equiv\ln(\exp(z) - 1),$ which we have written as $U(z)$ in the main body of the paper and plotted on the axes of Figure~\ref{fig:mah_pdf_bimodality}, as well as on the axes of the figures in Appendix~\ref{appendix:fitting_population_growth}. Figure~\ref{fig:softplus} gives a visual illustration of the behavior of the softplus function we use to enforce these constraints for the case of $\alate.$

Through our use of these variable transformations, we ensure that the best-fitting parameters determined by our application of gradient descent will always respect $0<\alate<\aearly.$  Mathematically, this guarantees that in the best-fitting approximation to each individual halo, for all $t>0$ the quantity $\mpeak(t)$ will increase monotonically, and $\dmhdt$ will strictly slow down as $\mpeak(t)$ eventually attains the value $\mzero$ at the time $t_0.$ These mathematical constraints embody the physical expectation that dark matter halo growth begins with a fast-accretion phase at early times, and transitions to a slow-accretion phase at late times. One can imagine employing a more flexible fitting function such as one based directly on a machine-learning algorithm that would not hard-wire this expectation into the approximation; such an approach has potential to capture short-term fluctuations that our model does not. We note, however, that even when these constraints are imposed on the fits as described above, dark matter halo growth can be approximated with a typical accuracy of 0.1dex or better for all $t\gtrsim1$ Gyr.

In carrying out this investigation, we studied a wide range of alternative implementations in which alternative variables were used in the optimization. For example, rather than using $\lge$ and $\lgl,$ we have also explored the use of logarithmic variables, and an alternative formulation in which the varied parameters were restricted to finite bounds via a sigmoid function. The specific choice for which parameters are varied in the optimization amounts, in effect, to a choice for a prior distribution on the parameters. We found through considerable experimentation that this choice can have a significant impact on the distribution of best-fitting parameters, even though the quality of the individual fits themselves was typically unaffected by this choice. Since one of our primary goals was to build a simple model for the distribution of best-fitting parameters, then the choice for the effective priors warranted special attention. Ultimately, our decision on the parameterization was driven by two considerations: first, that the differentiable approximation to each simulated halo respects the physical constraints defined above, and second, that the distribution of best-fitting parameters admits a reasonably simple empirical description.

Amongst all the variations we explored for the effective priors, as well as amongst several variations in the actual functional form used to fit the growth of individual halos, we found a clear bimodality in the halo population. The manifestation of this bimodality of course changed from variation to variation, but in all cases, we consistently found two separate sub-populations within the distribution of best-fitting parameters, roughly corresponding to early-forming and late-forming regions of parameter space. While these two sub-populations are easily identified in terms of their clustering within the distribution of best-fitting parameters, the distribution of the actual values of the formation time of the halos, $\tform,$ shows heavy overlap between the two groupings. There is also no strong dependence of the NFW concentration, $c,$ on sub-population membership, even though $c$ is strongly correlated with $\tform.$ Moreover, when splitting a sample of halos of the same $\mzero$ according to sub-population membership, the two-point clustering of two samples hardly differs, even for halo masses where the actual $\tform$-dependence of $\xi_{\rm m\delta}(r)$ is strong. Thus even though this bimodality considerably simplified our task of modeling the distribution of best-fitting parameters (as described in detail in Appendix~\ref{appendix:population_growth}), the sub-population membership of a halo may not have special physical significance.

\renewcommand{\thefigure}{B\arabic{figure}}
\section{Modeling the Assembly of Halo Populations with DiffmahPop}
\label{appendix:population_growth}
In this appendix, we describe our model for the diversity of assembly histories of halo populations. The goal of our halo population model  is to capture $\mahpdf$ and $\dmhdtpdf.$ Our approach to this problem is to use our model for individual halo growth as a basis, and to build a model of the statistical connection between present-day halo mass and the distribution of parameters describing individual halo MAHs, $\parampdf.$

As outlined in \S\ref{sec:halo_populations}, we model the distribution $\parampdf$ as a two-population normal distribution of 3 parameters. We model this distribution in terms of the same parameter transformations used in Appendix~\ref{appendix:individual_growth} to define the variables $\lgl\equiv U(\alate),$ $\lge\equiv U(\aearly-\alate),$ and $\lgtc\equiv\log_{10}\tauc,$ where $U(s)\equiv\ln(\exp(s) - 1)$ is the inverse of the softplus function (see Eq.~\ref{eq:softplus_constraints}). Thus each sub-population is described by an independently-defined mean, $\mui\equiv(\lge, \lgl, \lgtc),$ and covariance matrix, $\sigi,$ defined as
\beq
\sigi&\equiv&
  \begin{bmatrix}
   \sigma(\lge,\lge) & \sigma(\lge,\lgl) & \sigma(\lge,\lgtc)  \\
   \sigma(\lgl,\lge) & \sigma(\lgl,\lgl) & \sigma(\lgl,\lgtc)  \\
   \sigma(\lgtc,\lge) & \sigma(\lgtc,\lgl) & \sigma(\lgtc,\lgtc)
   \end{bmatrix}
\eeq
By defining our halo population model in terms of the transformed variables $\lge,\, \lgl,$ and $x_0,$ we ensure that all values of $\aearly,\, \alate,$ and $\tauc$ in the support of $\parampdf$ will respect the physical constraints described in \S\ref{subsec:constraints}, even in the tails of the distribution.

The relative weight of the two populations, $\flate,$ has shallow, monotonic mass-dependence, so that $\flate=\flate(\mzero).$ Each of the three components of $\mui,$ and each of the six components of $\sigi,$ also have a shallow, monotonic mass-dependence, so that $\mui=\mui(\mzero),$ and $\sigi=\sigi(\mzero).$  For the remainder of this appendix, we will describe our models for $\flate(\mzero),$ $\mui(\mzero)$ and $\sigi(\mzero)$ in turn.

We model the mass-dependence of $\flate(\mzero)$ as a power-law function of present-day mass with rolling index. We capture this behavior using the same sigmoid functional form defined in Eq.~\ref{eq:sigmoid}, so that $\flate$ has sigmoid-type dependence upon $\log_{10}\mzero.$ We calibrate the specific parameters of the sigmoid using the techniques described in detail in Appendix~\ref{appendix:fitting_population_growth}.

We model the mass-dependence of each component of $\mui$ in the same way we modeled $\flate,$ i.e., as a power-law function of $\mzero$ with rolling index, implemented based on a sigmoid function of $\log_{10}\mzero.$ Again we relegate discussion of our determination of the specific parameters of these sigmoid functions to Appendix~\ref{appendix:fitting_population_growth}.

In order to model the components of each $\sigi,$ we make use of the Cholesky matrix decomposition. Briefly, for any real-valued, symmetric, positive-definite matrix, $\Sigma,$ there exists a unique, real-valued, lower-triangular matrix, $L,$ that defines the Cholesky decomposition, written as $\Sigma=L\cdot L^{\rm T}.$ The diagonal entries of $L$ are strictly positive with a product equal to the determinant of $\Sigma.$ We use JAX to calculate $L$ so that our computations will be differentiable with autodiff, but many modern linear algebra libraries have efficient implementations of the Cholesky decomposition \citep[for a modern review, see][]{higham_2009_cholesky_review_article}.

In modeling each of the two $3\times3$ matrices, $\sigi,$ the quantities we parameterize are the 6 entries of the associated Cholesky matrix:
\beq
\label{eq:cholesky}
L_{\rm i}&\equiv&
  \begin{bmatrix}
   a_{\rm i} & 0 & 0  \\
   d_{\rm i} & b_{\rm i} & 0  \\
   e_{\rm i} & f_{\rm i} & c_{\rm i}
   \end{bmatrix}
\eeq
In fitting our model for $\mahpdf$ as described in Appendix~\ref{appendix:fitting_population_growth}, we formulate our parameterization in terms of $\log_{10}a_{\rm i},$ $\log_{10}b_{\rm i},$ and $\log_{10}c_{\rm i}$ to ensure that $\Sigma=L\cdot L^{\rm T}$ will always be positive definite, but the parameters $d_{\rm i}, e_{\rm i},$ and $f_{\rm i}$ can take on any value on the real line and still produce a valid covariance matrix, and so we use linear variables for our parameterization of the off-diagonal entries of $L.$

\renewcommand{\thefigure}{C\arabic{figure}}
\section{Fitting the Parameters of DiffmahPop}
\label{appendix:fitting_population_growth}
Our goal in \S\ref{sec:halo_populations} is to identify a point in the parameter space of our model that can be used to generate a realistic distribution of individual, smooth trajectories of halo assembly, and at the same time results in an accurate reproduction of $\mahpdf$ and $\dmhdtpdf.$ To achieve these goals, our model for $\mahpdf$ has numerous free parameters that required tuning in order to achieve the level of agreement shown in Figures~\ref{fig:mah_pdf_validation} \& \ref{fig:zform}. In \ref{appendix:differentiable_populations}, we review the general techniques we used for making differentiable predictions for the mean and variance of $\mahpdf$ and $\dmhdtpdf,$ and in \ref{appendix:population_optimization} we describe how we optimized the parameters of our best-fitting models.

\subsection{Differentiable predictions for halo population assembly}
\label{appendix:differentiable_populations}

The first moment of our model prediction for $\mahpdf$ is given by
\beq
\label{eq:mean_mah_pdf}
\langle\mpeak(t)\vert\mzero\rangle&\equiv&\int\int\int {\rm d}\tauc {\rm d}\aearly {\rm d}\alate\mpeak(t)\nonumber\\
&\times&\parampdf,
\eeq
where the PDF in the integrand is given by\footnote{Note that in Eq.~\ref{eq:mean_mah_pdf} we have written $\mpeak(t\vert\aearly,\alate,\tauc)$ simply as $\mpeak(t),$ and in Eq.~\ref{eq:composite_pdf} we have suppressed the conditioning upon $\mzero,$ e.g., $\flate(\mzero)$ is written simply as $\flate.$}
\beq
\label{eq:composite_pdf}
P(\aearly, \alate, \tauc) &=& (1-\flate)\cdot P_{\rm early}(\aearly, \alate, \tauc)\nonumber\\
&+&\flate\cdot P_{\rm late}(\aearly,\alate, \tauc),
\eeq
where $P_{\rm early}$ and $P_{\rm late}$ are the two, separately-defined three-dimensional Gaussian distributions described in Appendix~\ref{appendix:population_growth}. The calculation of the second moment of $\mahpdf$ is similar to Eq.~\ref{eq:mean_mah_pdf}, only rather than $\mpeak(t)$ appearing in the integrand, the PDF-weighted quantity is instead $\left[\mpeak(t)-\langle\mpeak(t)\rangle\right]^{2}.$

There are a number of approaches one could take to computing these nested integrals. First, we note that our model for $\parampdf$ is completely analytic and smooth, without any complex oscillatory behavior or sharp turns in the parameter space, and so commonly-used integration routines \citep[e.g.,][]{romberg_integration_1955} would have no trouble giving a high-accuracy result. Alternatively, each of the component ingredients of our model for the PDF in the integrand of Eq.~\ref{eq:mean_mah_pdf} have well-established numerical algorithms for generating stochastic realizations of the distributions, and so it would be straightforward to use a Monte Carlo approach to compute these integrals via moments of randomly generated samples. Either calculation could also be conveniently conducted in python using implementations in the {\tt scipy} library. However, we have instead opted for a third method based on weighted sampling that simplifies gradient calculations with autodiff.

Our autodiff-friendly method for calculating the integrals in Eq.~\ref{eq:mean_mah_pdf} can be easily understood in terms of a simple toy example of a generic PDF convolution:
\beq
\label{eq:pdf_convolution}
y = \int_{-\infty}^{\infty}{\rm d}xP(x)f(x)\approx\sum_{x_i\in D} \delta x_iP(x_i)f(x_i),
\eeq
where we will think of the integrand as some smooth target function, $f(x),$ weighted by a probability distribution, $P(x),$ and where in the second equation we have replaced the continuous integration with a finite sampling of points spanning the domain of support, $D.$ In other words, to calculate the integral in Eq.~\ref{eq:pdf_convolution}, rather than using an iterative algorithm such as \citet{romberg_integration_1955} that is challenging to implement in an autodiff library, instead one can use simple Gaussian integration with a sufficiently dense sampling grid. This toy problem is of the same form as the problem at hand in the evaluation of Eq.~\ref{eq:mean_mah_pdf} for cases where $P(x)=P(x\vert\theta)$ and $f(x)=f(x\vert\theta),$ only with this formulation, we can readily see how simple it is to compute gradients of $y$ with respect to model parameters, $\theta:$
\beq
\label{eq:toy_pdf_convolution}
\partial y/\partial\theta = \sum_{x_i\in D} \delta x_i\frac{\partial}{\partial\theta}P(x_i\vert\theta)f(x_i\vert\theta),
\eeq
By implementing the model components described in Appendix~\ref{appendix:population_growth} in an autodiff library such as JAX, evaluating expressions such as \ref{eq:toy_pdf_convolution} is entirely straightforward and does not require laboriously working out the analytical result for each contribution to the summations. One need only ensure that the integration domain $D$ is sufficient to span the range of non-negligible support for $\parampdf.$ We found this flexibility especially helpful during development phases of our model, when the need to conduct a range of modeling experiments required many variations of the underlying functional forms. In the following section, we describe how we carried out these calculations for the particular case of the PDF convolutions that arise from optimizing our model predictions for the first and second moments of $\mahpdf$ and $\dmhdtpdf.$

\subsection{Optimizing \dmahpop predictions for halo population assembly}
\label{appendix:population_optimization}

\begin{figure*}
\begin{centering}
\includegraphics[width=14cm]{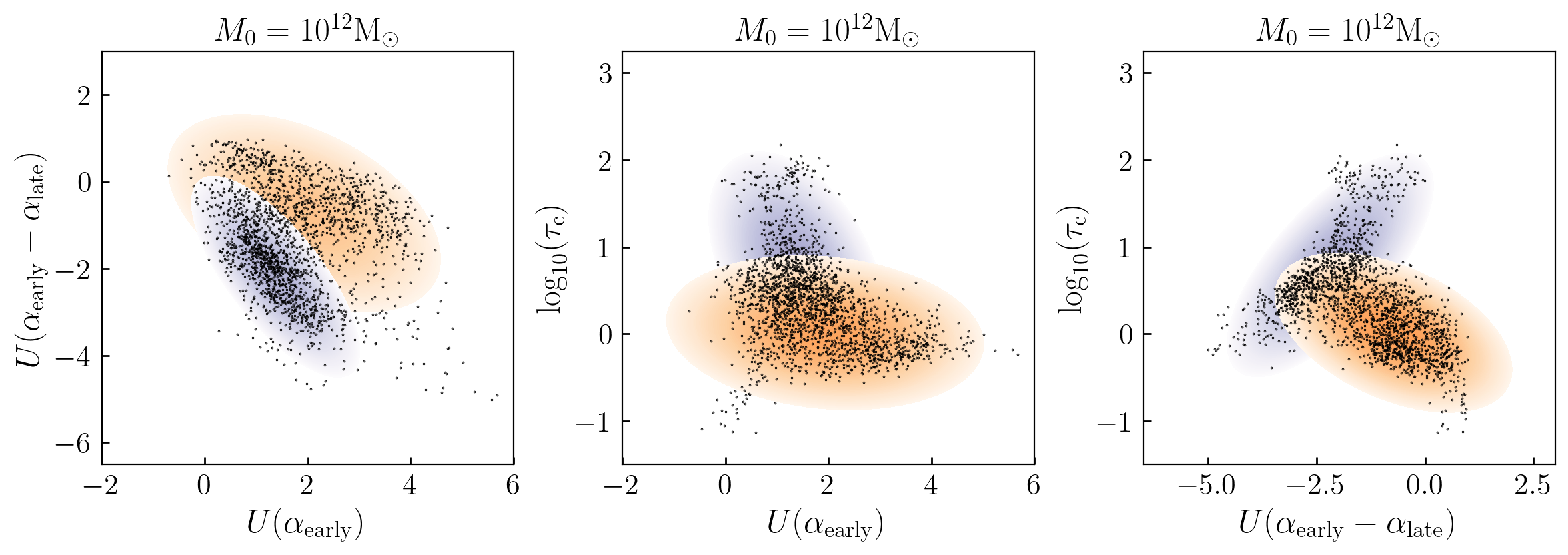}
\caption{{\bf Example comparison of the distribution of \dmah parameters for individual halo assembly}. For halos of present-day peak mass $\mzero=10^{12}\msun,$ the figure shows cross-sections of $\parampdf.$ For each individual BPL halo in the mass bin, we approximate the assembly history of the halo with the rolling power-law model described in \S\ref{sec:individual_halos}, and plot the distribution of parameters $U(\aearly),$ $U(\aearly-\alate),$ and $\log_{10}\tauc,$ where $U(s)\equiv\ln(\exp(s) - 1).$ The orange ellipsoids show the earlier-forming sub-population $(\tauc\lesssim2$ Gyr) of the two-component Gaussian distributions identified by {\tt scikit-learn}, and the purple ellipses show the later-forming sub-population $(\tauc\gtrsim2$ Gyr). The colored ellipses are defined by $\mubari$ and $\sigbari$ that we use to define the initial guess for the parameters regulating our model prediction for $\mahpdf$ and $\dmhdtpdf.$ See text in \ref{appendix:population_optimization} for details.}
\label{fig:mah_pdf_params_logmp12}
\end{centering}
\end{figure*}

\begin{figure*}
\begin{centering}
\includegraphics[width=13cm]{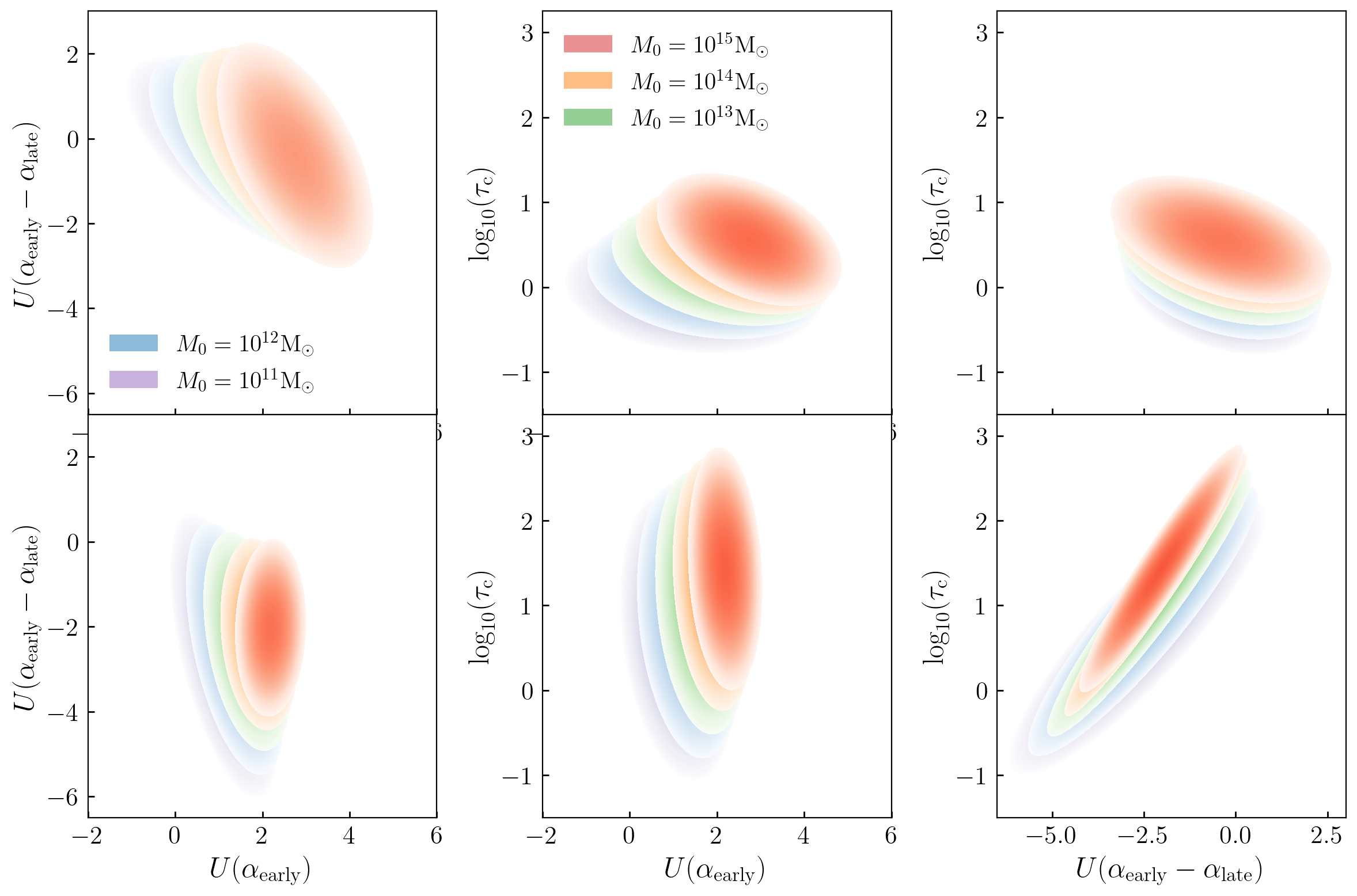}
\caption{{\bf Model for the mass-dependence of the distribution of the \dmah parameters for individual halo assembly}.  The axes in each row of panels are the same as those described in Figure~\ref{fig:mah_pdf_params_logmp12}. Different colored ellipses correspond to normal distributions of halos with different present-day mass as indicated in the legend. The top row (bottom row) of panels shows distributions for the early-forming (later-forming) population of halos, where the mean and variance of each population was determined according to the optimization procedure described in \ref{appendix:population_optimization}.}
\label{fig:mah_pdf_params_rockstar_massdep}
\end{centering}
\end{figure*}

In this section, we describe how we optimized the parameters of our model predictions for $\mahpdf$ and $\dmhdtpdf,$ where the target distributions come from the merger trees of halos in gravity-only N-body simulations. To generate our target data for a particular bin of present-day halo mass, we begin by selecting host halos in bin of $\mpeak(z=0)$ with width $0.1$dex. One issue that arises in making accurate predictions for the variance of halo histories is that in our model, the trajectory of each halo has exactly the same final mass, $\mzero,$ by construction, whereas the bin of simulated halos will have a source of unwanted variance due to the finite width of the bin. To account for this effect, we rescale the history of each simulated halo according to its actual value of $\mpeak(z=0)$ in the simulation before computing the target mean and variance. 

As mentioned in \S\ref{sec:halo_populations}, when defining the target data for a particular bin of present-day halo mass, $M_0,$ we limit the impact of wild outliers in the merger trees by trimming $3\sigma$ outliers from the halo sample. We conduct this outlier exclusion procedure as follows. For each snapshot of the simulation, $t,$ we compute the trimmed mean, $\mu_{\rm p}(t\vert M_0),$ and trimmed variance, $\sigma_{\rm p}(t\vert M_0),$ using {\tt scipy.stats.mstats.trimmed\_mean} and {\tt scipy.stats.mstats.trimmed\_variance}, respectively; within the {\tt scipy} library, these quantities are computed by rank-ordering the objects in the sample, computing the percentile of each object, $p,$ excluding the objects lying outside the range $(p, 100-p),$ and computing the mean and variance from the resulting population. At each snapshot, and for each bin of halo mass, we use $\mu_{\rm p}$ and $\sigma_{\rm p}$ to compute the $z$-score, $$z_{\rm p}(t\vert M_0)\equiv\frac{\log_{10}\mpeak(t\vert M_0)-\mu_{\rm p}(t\vert M_0)}{\sigma_{\rm p}(t\vert M_0)}.$$ We use $p=0.01$ to define each snapshot's trimmed mean and variance, and at each snapshot over the redshift range $0<z<3,$ we flag any halo with $\vert z_{\rm p}(t\vert M_0)\vert>3.$ We exclude any flagged halo from the sample whose mean and variance define the target data we use to optimize our model for halo populations; this outlier exclusion procedure typically removes 2-3\% of the halos in each mass bin; by construction, this excludes halos that experience an unusually large major merger after $z<3;$ the function implementing this algorithm can be found in the {\tt measure\_mahs.py} module of the publicly available {\tt diffmah} source code. To compute the target data defined in this fashion, we use the BPL simulation for bins of present-day halo mass centered at $\mzero\leq10^{13.5}\msun,$ and MDPL2 for larger halo masses, choosing bin centers spanning the range $\log_{10}\mzero\in\left[11.75,14.5\right],$ with 0.25dex separation.

In building our model for $\mahpdf,$ our goal is to faithfully recover the diversity of {\em smooth} assembly histories for a population of halos with the same present-day mass, $\mzero.$ As our model for individual halo assembly does not explicitly account for contributions from mergers, for our target data we use each individual halo's best-fitting $\mpeak(t)$ and $\dmhdtt$ to define the target mean and variance of $\mahpdf$ and $\dmhdtpdf.$ If we were to instead use the directly simulated histories to define the target one-point functions, then our model for $\parampdf$ would instead converge to a result with unrealistically large variance in the smooth trajectories traced by real halos. Using the directly simulated values of the MAHs as target data would require building an additional model component describing the variance from mergers about the smooth trajectories, which is beyond our intended scope.

The predictions of our model are controlled by the parametrized behavior of $\flate(\mzero),$ $\mui(\mzero)$ and $\sigi(\mzero),$ and optimizing these functions requires beginning with an initial guess for the parameters. In order to determine this initial guess, we directly inspect the distribution of best-fitting parameters describing the individual trajectories of the halos in each target bin, $\mzero.$ We use the Gaussian mixture model implemented in {\tt scikit-learn} to decompose the binned halos into two weighted sub-populations defined by $\lge, \lgl,$ and $\lgtc;$ we thus use {\tt scikit-learn} to supply an estimate for the fraction of later-forming halos in the bin, $\fbarlate(\mzero),$ and the mean and covariance of each sub-population, $\mubari(\mzero)$ and $\sigbari(\mzero),$ respectively. We proceed in this fashion and record each result for each bin of $\mzero.$ Figure~\ref{fig:mah_pdf_params_logmp12} shows an example comparison between initial guess for the two-component model and the true distribution of parameters for halos with $\mzero=10^{12}\msun.$

Using the collection of $\fbarlate(\mzero),$ $\mubari(\mzero)$ and $\sigbari(\mzero)$ for each of our target bins of present-day halo mass, we find that each of these quantities exhibits a shallow, monotonic mass-dependence that is well-approximated by the 4-parameter sigmoid function defined in Eq.~\ref{eq:sigmoid}. To define our initial guess for the best-fitting parameters of our halo population model, the parameters of each sigmoid can simply be hand-tuned to give a reasonable approximation to each quantity, and in all cases we found acceptable approximations by holding the transition-mass parameter fixed to $x_0=13.5,$ and the transition-speed parameter fixed to $k=0.5.$

Using the procedure described above to define the initial guess, we use the L-BFGS-B algorithm implemented in {\tt scipy} to optimize our model parameters. The loss function we minimize in this case is analogous to $\mathcal{L}_{\rm MSE}$ used in Appendix~\ref{appendix:individual_growth} (see Eq.~\ref{eq:mse_loss}), only here we minimize the sum of the logarithmic differences between the collection of means and variances of $\mahpdf$ and $\dmhdtpdf$ tabulated for each of our target mass bins. In searching for the best-fitting point, we do not vary all four parameters of each sigmoid function of each component, but only the two parameters controlling the two asymptotic bounds on each sigmoid, holding the $k$ and $x_0$ parameters fixed to their hand-tuned values. Moreover, we do not allow the parameters to vary arbitrarily, but rather we only permit the L-BFGS-B algorithm to search within a relatively narrow range of $\sim30\%$ of the initial guess. We refer the reader to our publicly available source code for further quotidian details on the implementation of this optimization procedure. Figure~\ref{fig:mah_pdf_params_rockstar_massdep} displays the results for the mass-dependence of the two-component normal distribution whose predictions for $\mahpdf$ and $\dmhdtpdf$ are shown in the main body of the paper. We refer the reader to the {\tt rockstar\_pdf\_model.py} of our source code for the specific best-fitting values that result from this optimization exercise.

\renewcommand{\thefigure}{D\arabic{figure}}
\section{Assembly Histories of Subhalos, Orphans, and Splashback-Centrals}
\label{appendix:subs}

\begin{figure*}
\begin{centering}
\includegraphics[width=13cm]{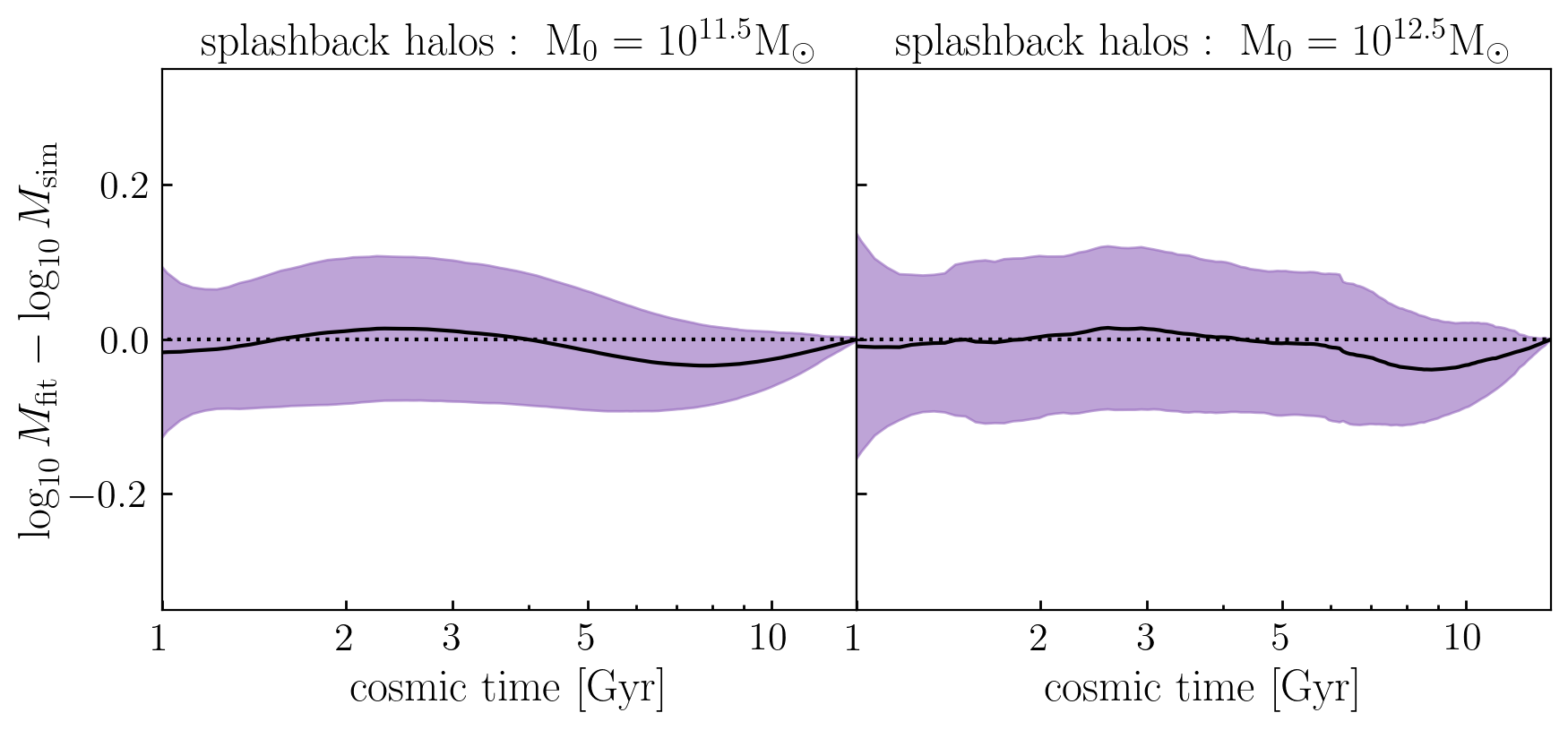}
\caption{Same as Fig.~\ref{fig:individual_residuals}, but for host halos in BPL that experienced a flyby event in their past history.}
\label{fig:splashback_residuals}
\end{centering}
\end{figure*}

\begin{figure*}
\begin{centering}
\includegraphics[width=13cm]{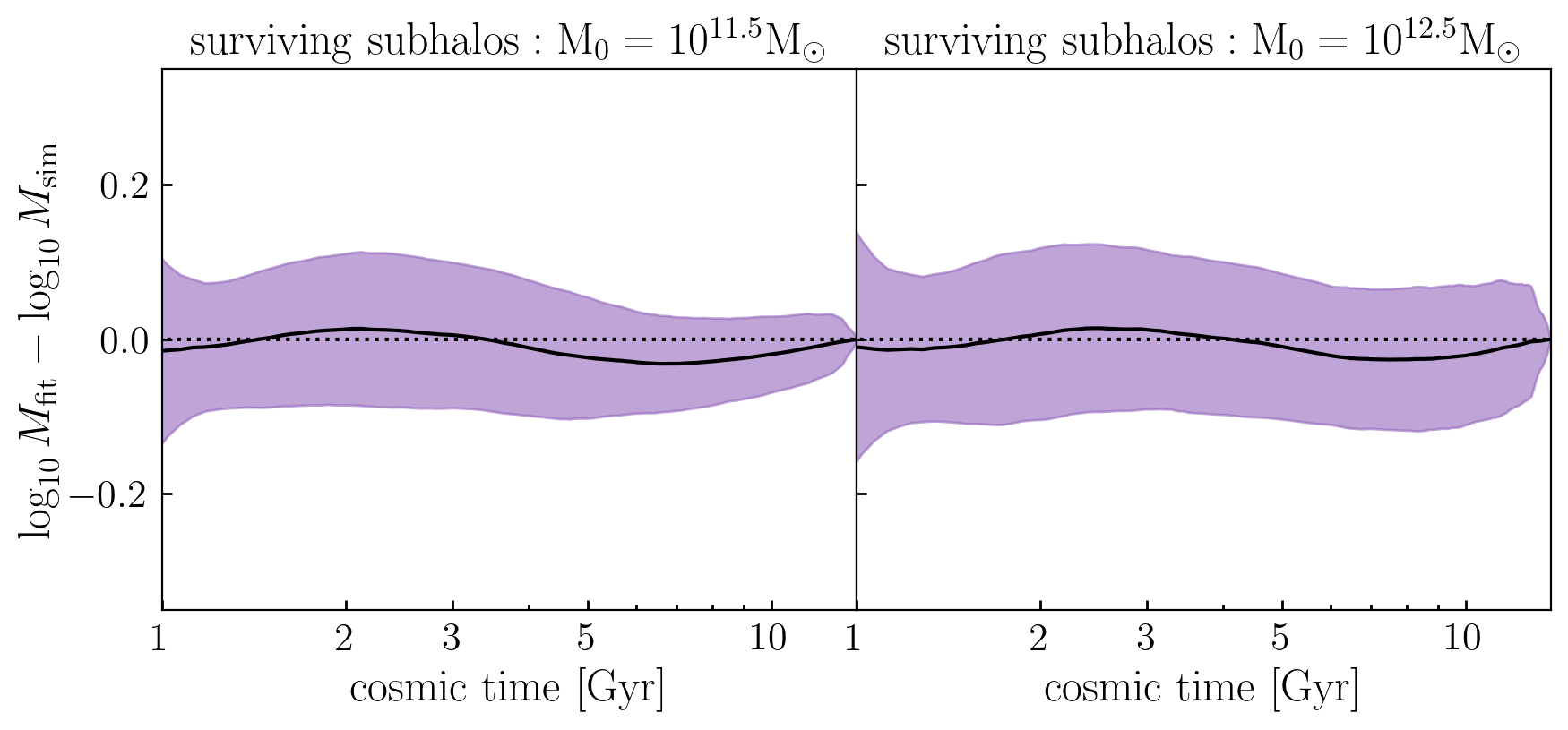}
\caption{Same as Fig.~\ref{fig:individual_residuals}, but for subhalos in BPL that survive to $z=0.$}
\label{fig:subhalo_residuals}
\end{centering}
\end{figure*}

\begin{figure*}
\begin{centering}
\includegraphics[width=13cm]{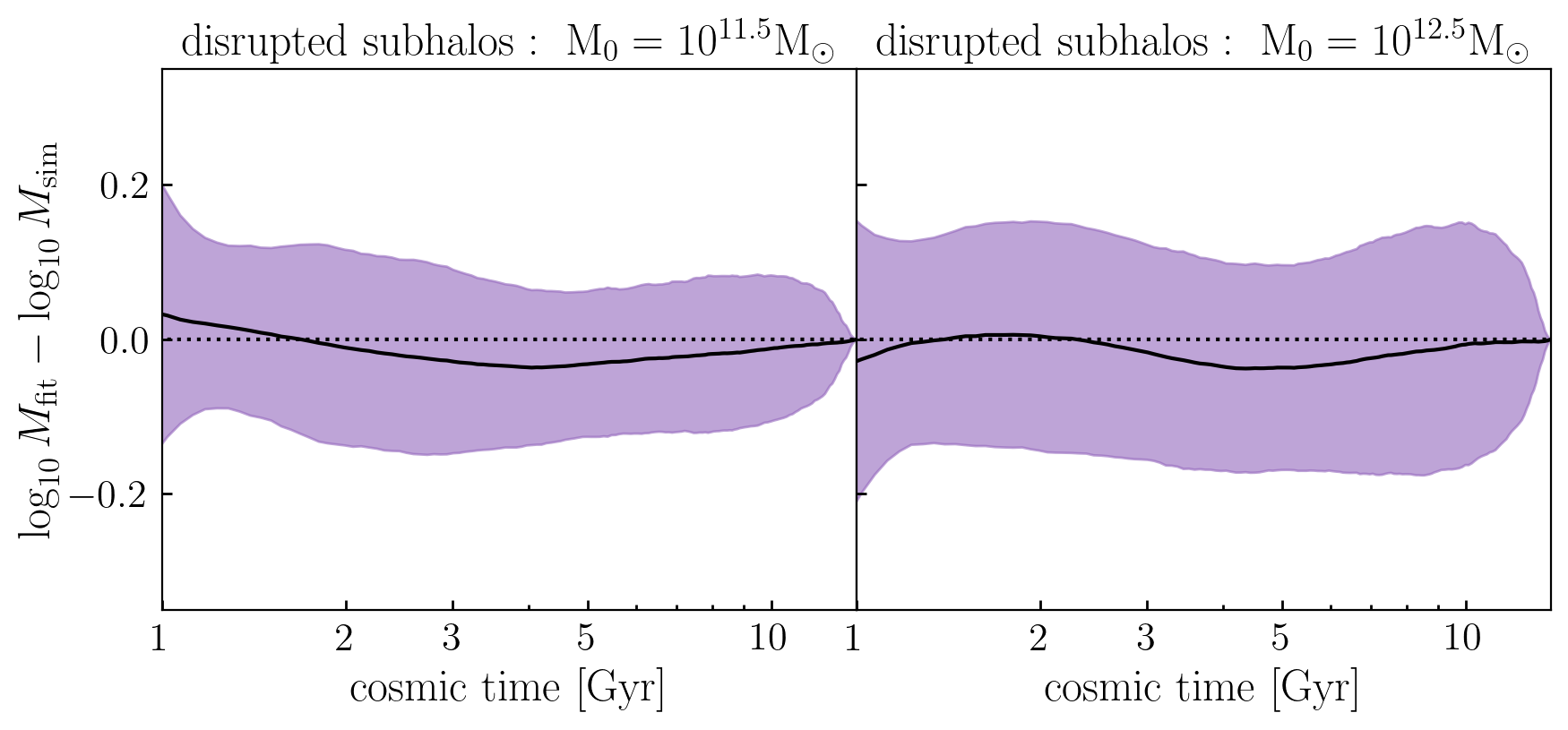}
\caption{Same as Fig.~\ref{fig:individual_residuals}, but for subhalos in BPL that either merged or were disrupted prior to $z=0.$}
\label{fig:orphan_residuals}
\end{centering}
\end{figure*}

In the publicly available Rockstar catalogs described in \S\ref{sec:sims}, the column {\tt a\_first\_infall} indicates the scale factor of the first snapshot where the (sub)halo was contained within the virial radius of some more massive halo. For a present-day host halo (as defined by {\tt upid=-1}), the value of {\tt a\_first\_infall} will be smaller than unity if and only if the  halo experienced a flyby event at some point in its past history. For the results in the main body of the paper, we defined our host halo samples by requiring  {\tt upid=-1}, without regard for the {\tt a\_first\_infall} column.  Figure~\ref{fig:splashback_residuals} in this appendix shows the residual errors of our model for individual growth for samples of flyby host halos defined by {\tt upid=-1} and {\tt a\_first\_infall<1}, using narrow bins of present-day peak halo mass as indicated by the in-panel annotation. Figure~\ref{fig:subhalo_residuals} in this appendix shows the analogous results for subhalos in the publicly available Rockstar catalogs in which {\tt upid}$\neq-1.$ Finally, Figure~\ref{fig:orphan_residuals} shows the analogous results for samples of subhalos that were either merged or disrupted at some point prior to $z=0.$ These subhalos do not appear in the publicly available Rockstar catalogs, but their histories can be extracted from the publicly available merger trees using the extract\_orphan\_info.c program in the UniverseMachine source code\footnote{https://bitbucket.org/pbehroozi/universemachine/src/main/}.

\renewcommand{\thefigure}{E\arabic{figure}}
\section{Assembly Histories of Halos in \tng}
\label{appendix:tng}

In this appendix, we illustrate the results of calculations described in the main body of the paper, only here shown for halos in \tng. When analyzing the \tng merger trees, we compute $\mpeak(t)$ using the {\em total} halo mass defined by the sum of dark matter, stars and gas. We remind the reader that halos and merger trees in \tng were identified by {\tt SUBFIND} and {\tt SUBLINK}, respectively, which are entirely different algorithms than the Rockstar and Consistent Trees codes used on the gravity-only simulations studied in this paper.

\begin{figure*}
\begin{centering}
\includegraphics[width=13cm]{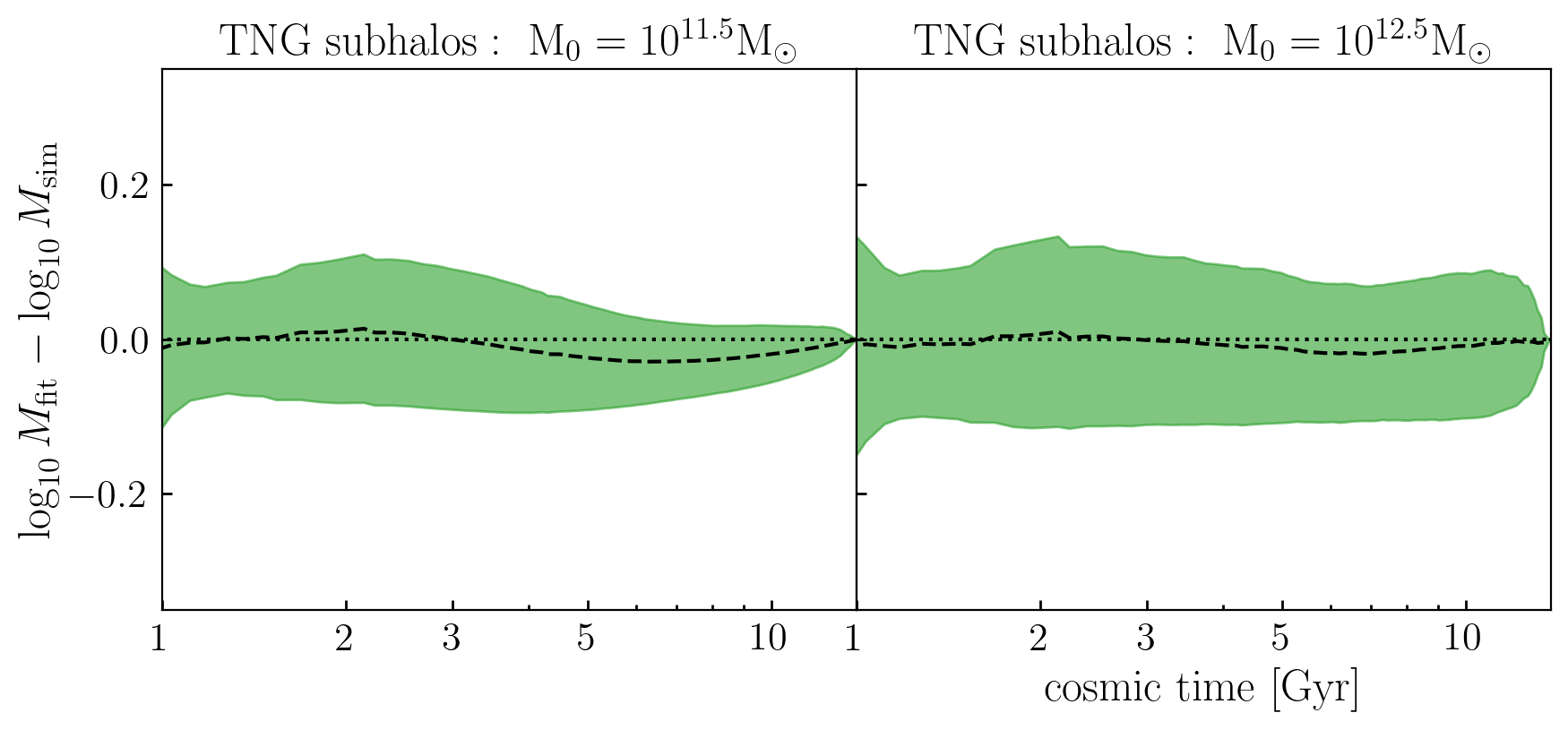}
\caption{Same as Figure~\ref{fig:individual_residuals}, but for subhalos in the \tng simulation.}
\label{fig:tng_subhalo_residuals}
\end{centering}
\end{figure*}

\begin{figure*}
\begin{centering}
\includegraphics[width=13cm]{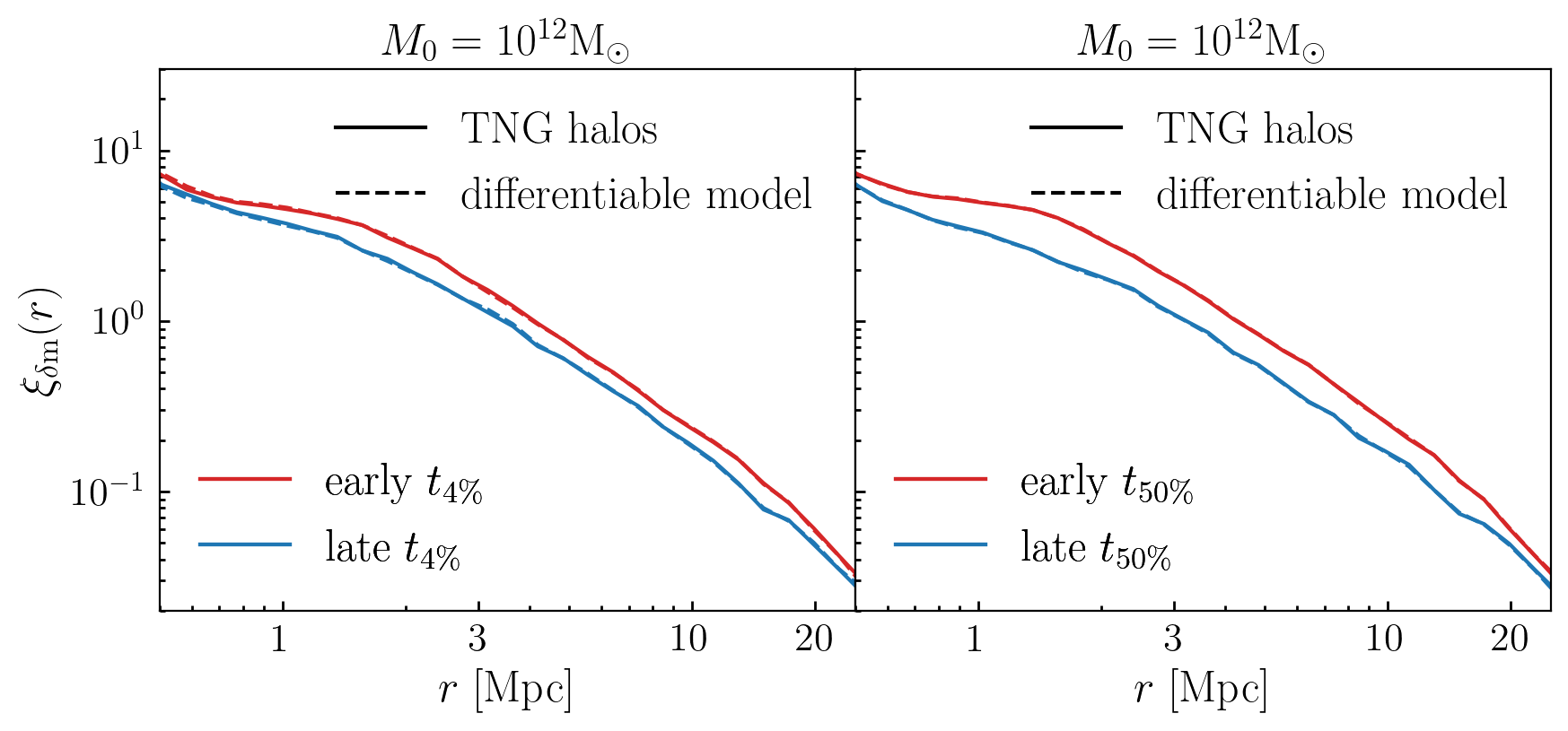}
\caption{Same as Figure~\ref{fig:assembias}, but for the \tng simulation.}
\label{fig:assembias_tng}
\end{centering}
\end{figure*}

\begin{figure*}
\begin{centering}
\includegraphics[width=13cm]{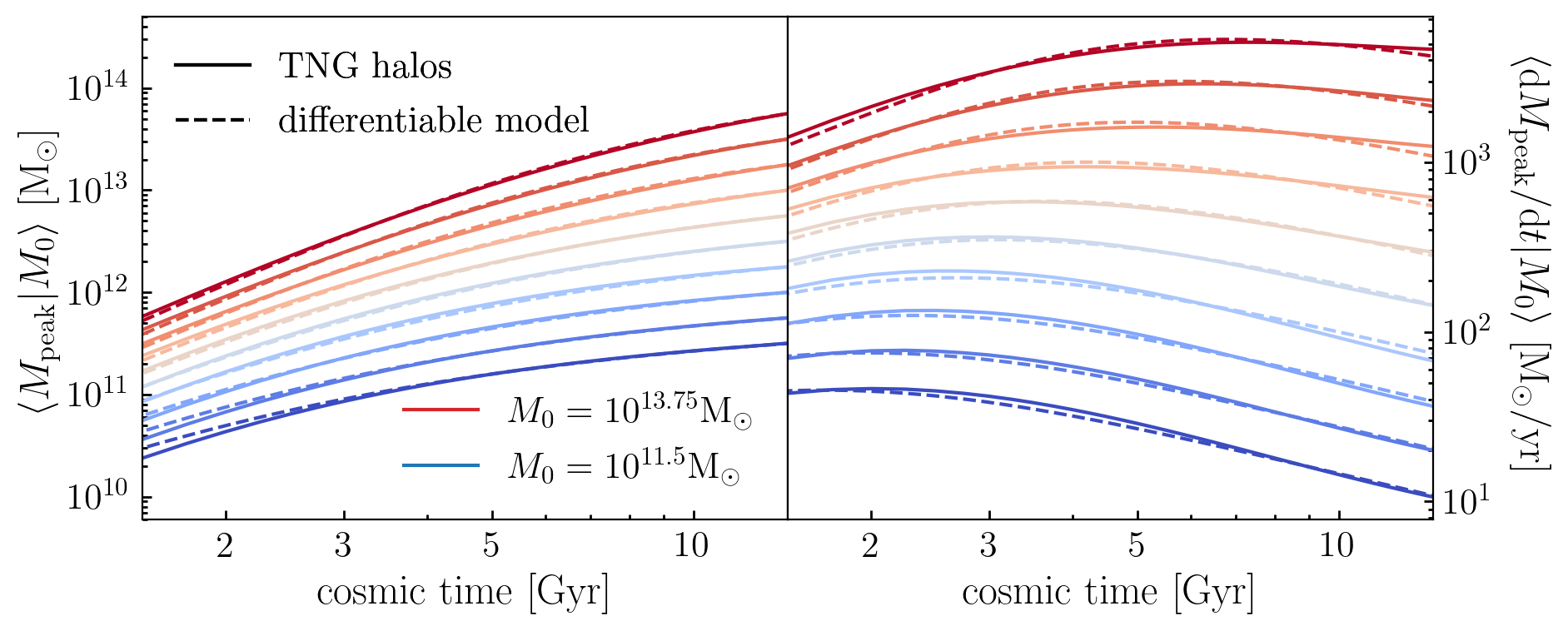}
\caption{Same as Figure~\ref{fig:avg_mah_validation}, but for the \tng simulation.}
\label{fig:tng_avg_mah_validation}
\end{centering}
\end{figure*}

\begin{figure*}
\begin{centering}
\includegraphics[width=14cm]{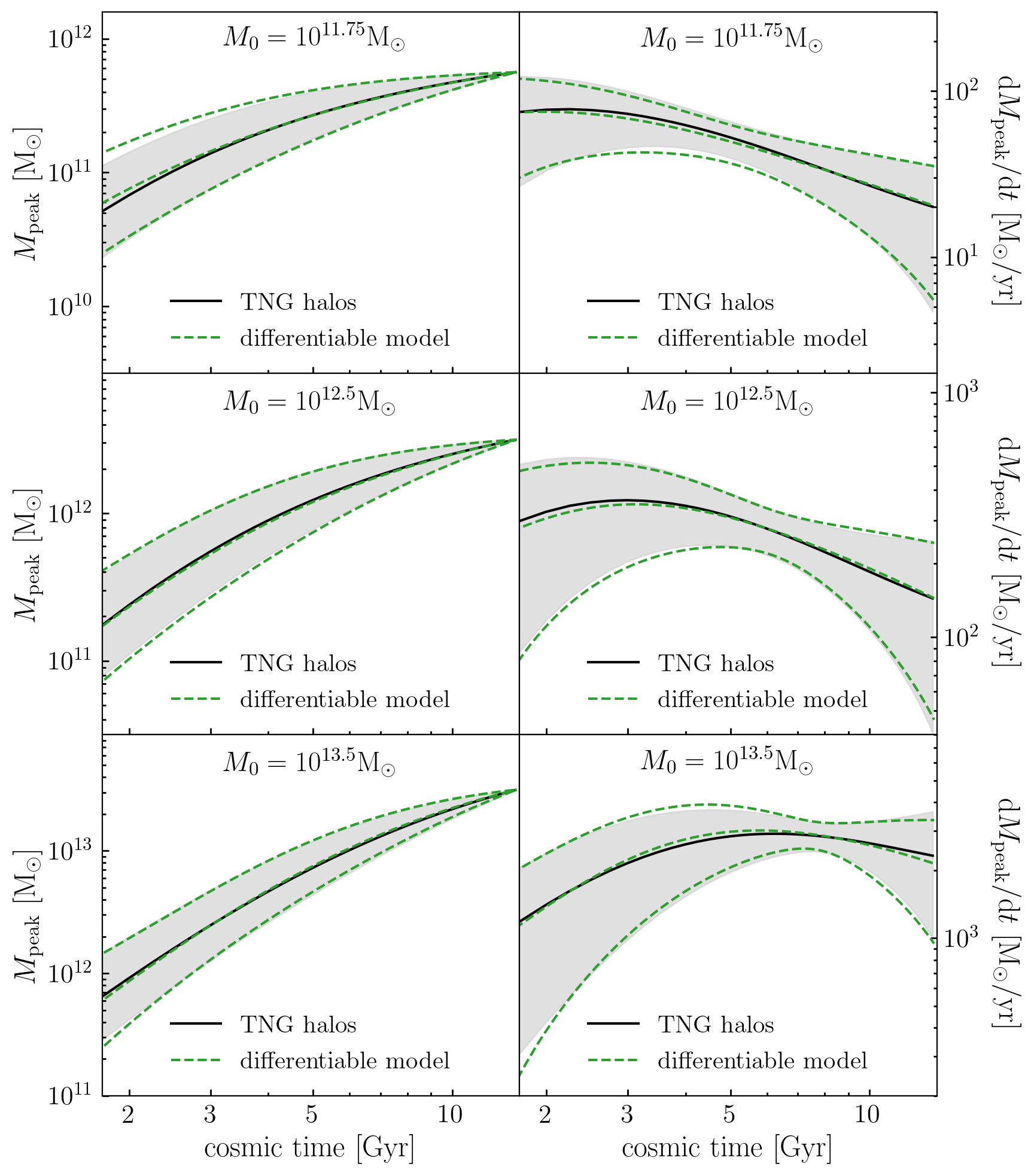}
\caption{Same as Figure~\ref{fig:mah_pdf_validation}, but for the \tng simulation.}
\label{fig:tng_mah_pdf_validation}
\end{centering}
\end{figure*}

Figure~\ref{fig:tng_subhalo_residuals} is analogous to Fig.~\ref{fig:individual_residuals} in the main body of the paper, but for the case of present-day subhalos in \tng. Figure~\ref{fig:assembias_tng} demonstrates that the residual errors in approximating halo growth are uncorrelated with the large-scale density field in \tng. For \tng, we have also repeated the halo population model calibration exercise described in detail in Appendix~\ref{appendix:fitting_population_growth}. Making no changes to the form of any model components, and varying the same set of parameters, we simply optimized the model according to a different set of target data. Figures~\ref{fig:tng_avg_mah_validation}-\ref{fig:tng_mah_pdf_validation} show that our model for halo population growth achieves a comparable level of accuracy for \tng as for gravity-only simulations.

\end{document}